\documentclass[fleqn,usenatbib]{mnras}
\usepackage{float}
\usepackage{amsmath,amssymb}
\usepackage{natbib}    
\usepackage{hyperref} 
\usepackage{graphicx} 
\usepackage{color}
\usepackage[verbose]{placeins}
\usepackage{stfloats}
\usepackage{verbatim}    
\usepackage{enumitem}
\usepackage{natbib}
\usepackage{CJK}
\usepackage{multirow}
\usepackage{comment}
\usepackage{bm}
\usepackage{physics}
\usepackage{array}
\usepackage{orcidlink}  

\newcolumntype{L}[1]{>{\raggedright\arraybackslash}p{#1}}
\newcolumntype{C}[1]{>{\centering\arraybackslash}p{#1}}
\newcolumntype{R}[1]{>{\raggedleft\arraybackslash}p{#1}}

\newcommand{\cntextsc}[1]
{\begin{CJK*}{UTF8}{gbsn}#1\end{CJK*}}
\newcommand{\proptosim}{\mathrel{\vcenter{
 \offinterlineskip\halign{\hfil$##$\cr
 \propto\cr\noalign{\kern2pt}\sim\cr\noalign{\kern-2pt}}}}}

\newcommand{\tp}{$\tau_p$}

\newcommand{\K}{\mathrm{K}}

\renewcommand{\P}{\mathrm{P}}     

\renewcommand{\d}{\mathrm{d}}

\renewcommand{\roman}[1]{\textsc{\MakeLowercase{#1}}}
\newcommand{\acc}{\mathrm{acc}}

\newcommand{\tctext}[1]{\begin{CJK}{UTF8}{bkai}#1\ignorespacesafterend\end{CJK}}

\usepackage{soul}

\graphicspath{{./}{figures/}}

\title[Gap Opening in MHD Protoplanetary Disks]
{3D Gap Opening in Non-Ideal MHD Protoplanetary Disks: Asymmetric Accretion, Meridional Vortices, and Observational Signatures}

\author[X. Hu et al.]{
Xiao Hu (\cntextsc{胡晓})$^{1,2}$\thanks{E-mail: xiao.hu.astro@gmail.com}\orcidlink{0000-0003-3201-4549},
Zhi-Yun Li$^{2}$,
Jaehan Bae$^{1}$\orcidlink{0000-0001-7258-770X},
Zhaohuan Zhu (\tctext{朱照寰})$^{3,4}$\orcidlink{0000-0003-3616-6822}
\\
$^{1}$Department of Astronomy, University of Florida, Gainesville, FL 32608, USA\\
$^{2}$Department of Astronomy, University of Virginia, Charlottesville, VA 22904, USA\\
$^{3}$Department of Physics and Astronomy, University of
  Nevada, Las Vegas, 4505 S. Maryland Parkway, Las Vegas,
  NV 89154, USA\\
$^{4}$Nevada Center for Astrophysics, University of Nevada, Las Vegas, 4505 South Maryland Parkway, Las Vegas, NV 89154, USA}

\date{Accepted XXX. Received YYY; in original form ZZZ}

\pubyear{2023}

\begin{document}
\label{firstpage}
\pagerange{\pageref{firstpage}--\pageref{lastpage}}
\maketitle

\begin{abstract}
Recent high-angular resolution ALMA observations have revealed rich information about protoplanetary disks, including ubiquitous substructures and three-dimensional gas kinematics at different emission layers. One interpretation of these observations is embedded planets. Previous 3-D planet-disk interaction studies are either based on viscous simulations, or non-ideal magnetohydrodynamics (MHD) simulations with simple prescribed magnetic diffusivities. This study investigates the dynamics of gap formation in 3-D non-ideal MHD disks using non-ideal MHD coefficients from the look-up table that is self-consistently calculated based on the thermo-chemical code. We find a concentration of the poloidal magnetic flux in the planet-opened gap (in agreement with previous work) and enhanced field-matter coupling due to gas depletion, which together enable efficient magnetic braking of the gap material, driving a fast accretion layer significantly displaced from the disk midplane. The fast accretion helps deplete the gap further and is expected to negatively impact the planet growth. It also affects the corotation torque by shrinking the region of horseshoe orbits on the trailing side of the planet. Together with the magnetically driven disk wind, the fast accretion layer generates a large, persistent meridional vortex in the gap, which breaks the mirror symmetry of gas kinematics between the top and bottom disk surfaces. Finally, by studying the kinematics at the emission surfaces, we discuss the implications of planets in realistic non-ideal MHD disks on kinematics observations.
\end{abstract}

\begin{keywords}
accretion, accretion disks ---
  magnetohydrodynamics (MHD) --- planets and satellites:
  formation --- circumstellar matter --- method: numerical
\end{keywords}



\section{Introduction} 
\label{sec:intro}

Recent observations indicate that protoplanetary disks ubiquitously exhibit substructures \citep[e.g.,][]{2015ApJ...808L...3A,2018ApJ...869L..42H,2020ARA&A..58..483A}, including rings and gaps, which are present in both gas and dust components \citep{2023ASPC..534..423B}. Numerous theoretical models have been developed to explain these features. Inherent physicochemical processes within the disk, such as the sublimation fronts of different volatile, i.e., snow lines, can influence dust growth and thereby lead to the formation of rings and gaps \citep{2015ApJ...806L...7Z,2016ApJ...821...82O}. Interactions between dust and gas, such as direct aerodynamic drag \citep[e.g.,][]{2017MNRAS.467.1984G}, and indirect effects from non-ideal MHD processes \citep{2019ApJ...885...36H,2021ApJ...913..133H}, can also affect local accretion rates, giving rise to substructures. On the other hand, a popular explanation posits that these rings and gaps are indicative of young planets in formation \citep[see the PPvii review by][]{2023ASPC..534..423B}. 

High-resolution observations of molecular line emissions have uncovered more detailed gas structures within protoplanetary disks \citep[e.g.,][]{2019Natur.574..378T,2020ApJ...890L...9P}. Encouragingly, the substructures' azimuthal velocities closely match the perturbations caused by planets, supporting the increasingly accepted hypothesis of gap opening by massive planets. On the other hand, radial and vertical velocities exhibit considerably complex structures. The Molecules with ALMA at Planet-forming Scales (MAPS) \citep{2021ApJS..257....1O} survey conducted spatial mapping of multiple chemical species across five protoplanetary disks: MWC 480, HD 163296, AS 209, IM Lup, and GM Aur. Notably, the vertical velocities near the gaps in HD163296 and AS209 \citep{2023A&A...674A.113I,2023ApJ...950..147G} do not align with the collapsing flows predicted by hydrodynamic simulations \citep{2001ApJ...547..457K, 2016ApJ...832..105F}, casting doubts on the planet-disk interaction interpretation.

Gap opening in protoplanetary disks by planets is a classic problem in planet formation, wherein the planet's gravity launches spiral density waves, depositing significant torque in the disk \citep[e.g.,][]{1993prpl.conf..749L}. This torque alters the disk structure, resulting in a local low-density annulus (i.e., a planetary gap) along the planetary orbit, influencing gas and solids accretion as well as orbital migration of the planet \citep[e.g.,][]{2006Icar..181..587C,2014prpl.conf..667B}. While early studies have been carried out adopting hydrodynamic, viscous disks, recent studies have shown that the accretion and angular momentum transport in protoplanetary disks are governed by magnetically-driven disk winds \citep[e.g.,][]{2017A&A...600A..75B, 2019ApJ...874...90W}. Some previous hydrodynamical simulations have introduced wind torque to approximate the effects of wind-driven accretion in disks. Even in simplified one-dimensional models, such as those by \citet{2015A&A...584L...1O,2017A&A...608A..74O}, quite intriguing results have been found regarding the migration of low-mass planets: the inward migrational torque is weakened, and can even be reversed. In two-dimensional simulations, \citet{2020A&A...633A...4K} found that the migration of massive planets could also reverse under certain conditions. \citet{2022A&A...658A..32L} and \citet{2022MNRAS.515.3113E} pointed out more complex behaviors, such as the ease of vortex formation at the edge of the gap by wind-driven accretion, and a reduction in the timescale for gap opening, both of which can influence the rate of planet migration and the criteria for gap opening mass. \citet{2022MNRAS.512.2290T} employ a parameter similar to the viscous $\alpha$, denoted as $\alpha_{dw}$, to evaluate the wind loss rate in one-dimensional model, facilitating the study of the long-term evolution of wind-driven accretion disks.

While these hydro models have made significant progress using simplified disk wind treatment, it is widely recognized that magnetic fields play a crucial role in both launching disk winds and the corresponding angular momentum transfer processes. Magnetic fields have been shown to play important roles in carving gaps by direct surface accretion stream \citep{2017MNRAS.468.3850S} or radial magnetic flux redistribution \citep[e.g.,][]{2018MNRAS.477.1239S,2019ApJ...885...36H, 2021ApJ...913..133H,2021MNRAS.507.1106C}. In turn, the spiral wakes and gaps generated by planets should also affect the distribution of magnetic fields. The wind torque and wind loss rate are intricately linked to the disk's magnetization, determined by its surface density and the strength of the poloidal magnetic field \citep{2020A&A...639A..95R,2021A&A...650A..35L}. 
Their dynamical importance motivated numerical studies that combine planet-disk interaction with magnetic fields
and wind-driven accretion. Early simulations along this line usually assumed ideal MHD and incorporated only a toroidal magnetic field without launching a disk wind \citep{2003MNRAS.339..993N,2003ApJ...589..543W,2011A&A...533A..84B}. More recent local shearing-box MHD simulations have included a net poloidal magnetic flux and found that the magnetic flux
gets concentrated into the planet-induced gap, making
the gap deeper and wider due to enhanced magneto-rotational instability (MRI) turbulence within the gap \citep{2013ApJ...768..143Z,2017MNRAS.472.3277C}. However, besides being local, these studies are typically unstratified in the vertical direction and thus incapable of wind-driven angular momentum transport.

Once magnetic fields are considered, non-ideal MHD processes become crucial, as outer protoplanetary disks are weakly ionized. \citet{2013ApJ...779...59G} included Ohmic dissipation in a global simulation, but its extent is limited to 4.5 disk scale heights, limiting the disk wind's treatment. 
\citet{2023ApJ...946....5A} conducted three-dimensional global non-ideal magneto-hydrodynamic simulations with ambipolar diffusion to investigate Type-II planet-disk interactions. Their findings highlighted how embedded planets could lead to a concentration of poloidal magnetic flux around their orbits, enhancing angular momentum removal and deepening the planetary gaps formed. This work emphasizes the crucial influence of magnetic fields and MHD winds on gap formation and planetary migration. 
Similarly, \citet{2023A&A...677A..70W} explored the dynamics of planet-disk interactions through high-resolution 3D global non-ideal MHD simulations, examining the effects of various planet masses and disk magnetizations on gap opening and meridional flows. Their research pointed to the significant impact of MHD winds on shaping disk structures and influencing the migration of embedded planets, reinforcing the importance of incorporating magnetic fields into models of planet formation and disk evolution.

However, these studies employed a simplified approach to magnetic diffusion, using a fixed profile for the ambipolar Elsasser number (Am), which may not capture the full complexity of magnetic interactions within protoplanetary disks. 
As the gas density changes due to the development of MHD winds, the ionization structure of the disk also changes which in turn affects the efficiency of the coupling between the magnetic fields and the gas. It is thus expected that the dimensionless Elsasser number, characterizing the diffusion timescale normalized by the local orbital period in protoplanetary disks would have spatial and temporal variations. Capturing this variation more self-consistently based on the recent thermochemical calculation (including ionization) of \citet{2023MNRAS.523.4883H} is a major goal of our investigation. 

The paper is organized as follows. In \S\ref{sec:methods}, we describe the numerical methods and simulation setup. \S\ref{sec:model-fiducial} analyzes the results from our simulation, focusing on the gas dynamics and the corresponding observational signatures. We discuss our results and conclude in \S\ref{sec:discussion}.

\section{Methods}
\label{sec:methods}
\subsection{Disk}

We simulate the non-ideal MHD disk evolution using \verb|Athena++| \citep{2020ApJS..249....4S}. To minimize the grid noise of Keplerian rotation and the radially flowing disk wind, we perform simulation in spherical polar coordinates $(r,\theta,\phi)$. The radial grid in our simulation is set from r = 2 to 100 in code units with logarithmic grid spacing. The $\theta$ grid extends from 0.05 to $\pi-0.05$. The azimuthal $\phi$ grid spans from $0$ to $2\pi$. The root grid has 64, 48, and 96 cells in $r$, $\theta$, and $\phi$ directions, respectively. We used 4 levels of static mesh refinement. The first 2 levels are axisymmetric, covering the whole gap region, within $4<r<23$ and $\theta_{\rm mid}<0.5~{\rm rad}$, where $\theta_{\rm mid}$ is the angle above and below the midplane. The 3rd and 4th levels only cover the planet's vicinity, giving a resolution of 16 cells per scale height at the finest level. We use a corotating frame centered on the planet \citep{2023ApJ...946....5A} so the static mesh refinement can follow the circumplanetary disk region.

The disk midplane's initial density and temperature profile follow power-law functions with indices of $p$ and $q$, respectively. The midplane density at $r=1$ is set to unity in code units.
The initial disk profile for density and temperature at the midplane is therefore:
\begin{eqnarray}
\rho(R,z=0)&=&\rho(R_0,z=0)\left(\frac{R}{R_0}\right)^p \label{eq:dslope}\\
T(R,z=0)&=&T(R_0,z=0)\left(\frac{R}{R_0}\right)^q \label{eq:tslope}
\end{eqnarray}
We adopt $p=-2.2218$ and $q=-0.57$ to match the disk profile in \citet{2023MNRAS.523.4883H}, and here $R=r\sin{\theta}$, $z=r\cos{\theta}$ are cylindrical coordinates. The entire temperature structure is adopted from a smooth disk (i.e., no-planet torque) run from \citet{2023MNRAS.523.4883H}, and the temperature for each cell is calculated using bilinear interpolation in the meridian plane. In general, the disk is vertically isothermal within four scale heights above the midplane. Beyond this, the temperature increases by 50\% within one additional scale height and then doubles at ten scale heights. The temperature of the corona continues to rise at higher altitudes, eventually plateauing at 2000-3000 K. This gives a midplane aspect ratio of 0.058 at R=10, and 0.035 at R=1, corresponding to T=82 K and 305 K when the code length unit equals 1 au. The disk is vertically isothermal with a hot corona. We use instant cooling to maintain a fixed temperature profile throughout the simulation.
For the vertical density structure, the initial profile is calculated by assuming hydrostatic equilibrium in the $R-z$ plane, i.e., $v_R=v_z=0$. As in \citet{2023MNRAS.523.4883H}, the disk is initially threaded by a large-scale poloidal magnetic field, with a midplane plasma~$\beta$ (defined as the ratio between thermal pressure and magnetic pressure $\beta\equiv2P_{gas}/B^2$) of $10^4$. The corresponding initial vector potential is adopted from \citet{2007A&A...469..811Z}:
\begin{equation}
A_\phi(r, \theta) = \frac{2B_{z0}R_0}{4+p+q}\left(\frac{r\sin\theta}{r_0}\right)^{\frac{p+q}{2}+1}\
[1+(m\tan\theta)^{-2}]^{-\frac{5}{8}}
\end{equation}
where $p$, $q$ and $r_0$ are from Eq.\ref{eq:dslope},\ref{eq:tslope}, and $m$ is a parameter that specifies the degree that poloidal fields bend, with $m\rightarrow\infty$ giving a pure vertical field. We chose $m=0.5$ the same as \citet{2017ApJ...836...46B}. We also applied the same disk setup to a hydro-only run, adopting an $\alpha=0.001$ and without the magnetic field.

\subsection{Planet}

The planet is held on a fixed circular orbit at $r_p$ = 10 code units (c.u. hereafter) in the disk midplane. The planet-to-primary mass ratio $q$ is set to 0.001, corresponding to the Jupiter mass around a Sun-like star, equivalent to 5.12 thermal mass. The total gravitational potential $\Phi$ considered in the simulation is $\Phi=\Phi_*+\Phi_p$, where $\Phi_*=-GM_*/r$ is the gravitational potential of the central star, and $\Phi_p$ is the potential of the planet, given by:
\begin{equation}
    \Phi_p=-\frac{GM_p}{\sqrt{|\bm{r}-\bm{r_p}|^2+r_s^2}}
    \label{eq:potential}
\end{equation}
where $M_p$ is the planet mass, $\bm{r}$ and $\bm{r_p}$ are vectors of
each grid cell center and of the planet from the central star, and
$r_s = 0.095$ is a smoothing length chosen to be 2.35 times the smallest cell size in the planet's vicinity. We ignore the indirect potential term that is due to the offset between the central
star and center of gravity of the star-planet system, as it has a negligible impact on the perturbation structure \citep{2019ApJ...875...37M}. We also ignore the planet's mass increase due to accretion since it is small during the simulation.  To minimize the influence of sudden planet insertion, the planet's mass is linearly increased with time, from zero at the start of the simulation to its final value at the time $t=2.3~\tau_p$ where $\tau_p$ is the planet's orbital period.

\subsection{Magnetic Diffusion}
\label{sec:diffusivity}

In a protoplanetary disk, we consider a weakly ionized fluid composed of ions, electrons, charged grains, and neutrals. 
We will only consider Ohmic dissipation and ambipolar diffusion, with the following induction equation governing the magnetic field evolution:  
\begin{equation}
\frac{\partial \bm{B}}{\partial t} = \bm{\nabla} \times \left(\bm{v} \times \bm{B}\right) - \frac{4\pi}{c}\nabla\times\left( \eta_O\bm{J}+\eta_A \bm{J_\perp} \right).
\label{eq:induction2}
\end{equation}
Here $\bm{J_\perp} = \bm{B}\times(\bm{J} \times \bm{B})/B^2$ is the current density perpendicular to the magnetic field, and $\eta_O$ and $\eta_A$ are the Ohmic and ambipolar diffusivities, respectively. 

The magnetic diffusivities, particularly $\eta_A$, depend strongly on the abundance and size distribution of dust grains, especially small grains, which can become the main charge carriers in dense regions of protoplanetary disks \citep{2019ApJ...874...90W}. Ideally, they should be computed through consistent thermochemistry simultaneously with the magnetohydrodynamics. However, the computation is prohibitively expensive, which motivates us to employ the look-up table\footnote{The table is available in both Python and C scripts at \url{https://github.com/astroxhu/diffusion-table}} from our precursor study where we adopted a 2D r-theta domain assuming axisymmetry  \citep{2023MNRAS.523.4883H}. The 2D model used a planetary torque to open a gap and tracked the evolution of magnetic diffusivities in the gap center to assemble the look-up table. In the full thermo-chemical setup, the disk has a midplane density of $8\times10^{-12}~{\rm g~cm^{-3}}$ at 10 au, which corresponds to a gas surface density of $170~{\rm g~cm^{-2}}$. 10 au is also the planet's distance to the central star, which equals 10 c.u. hereafter. There are 28 chemical species, including charged grains. Ionization sources include ray-tracing FUV and X-ray flux from the central star and diffused ionization from down-scattered X-ray, cosmic ray, and short-lived radioactive nuclei. We refer the reader to \citet{2023MNRAS.523.4883H} and references therein for a more detailed description. For the ambipolar diffusion, we obtain the Elsasser number $Am$ from the table; for the Ohmic resistivity, we directly obtain the diffusivity $\eta_O$ from the table. There's a simple monotonic trend between $\eta_O$ and the local density for the Ohmic resistivity. The $Am$ look-up table is a function of the local density normalized by the initial midplane value $\rho/\rho_{\rm mid,0}$ and the vertical location normalized by the scale height $z/h$. There are two motivations to use $\rho/\rho_{\rm mid,0}$ instead of the absolute value. First, we'd like to keep our 3-D setup as scale-free as possible. The original 2-D run used absolute density because of the chemical network. Second, the constant $Am\sim10^{-2}$ along the midplane is a good approximation for the full thermochemical model. The $Am$'s location dependence is within 2h, where the cosmic ray ionization is more attenuated from the column density. The $Am$ vertically varies only within 2h from the midplane, where the cosmic ray ionization is significantly attenuated due to the higher column density. Above that $Am$ only relies on $\rho/\rho_{\rm mid,0}$. There is an $Am$ ``valley" just below the disk surface, reaching $\sim$3h. This trend resembles the power-law prescription proposed by \citet{2018MNRAS.477.1239S} where the $Am$ inside the disk decreases with a decreasing density as a result of balancing the recombination of the dominant ions and electrons with the cosmic ray ionization, as discussed in \citet{2023MNRAS.523.4883H}.  

Since the look-up table is obtained in an environment where the radiation from the central star is sufficiently shielded, it only works well for the denser part of the disk, not the disk atmosphere. Thus, we keep the diffusion profile unchanged from the initial setup for any region more than 4$h$ from the midplane.

\subsection{Boundary Conditions}
\label{sec:boundary}
A modified outflow condition is employed for hydrodynamic quantities for the inner radial boundary. Here, the vector quantities' $r$ and $\theta$ components are duplicated from the last active grid cell into the ghost zones. We extrapolate the initial power-law distribution for the density $\rho$ (i.e., Eq.~\ref{eq:dslope}) and the Keplerian rotation for $v_\phi$ into the ghost zones. When directed toward the active simulation domain, the radial components of the vector quantities are set to zero in the ghost zones. The azimuthal magnetic field component ($B_\phi$) is set to zero at the inner radial boundary. At the outer radial boundary, all hydrodynamic and magnetic quantities are replicated from the active zones into the ghost zones, except when the radial velocity ($v_r$) is negative; in which case $v_r$ is set to zero in the ghost zones. Reflecting boundary conditions are applied at the $\theta$ boundaries adjacent to the polar axis.
A periodic boundary condition is utilized in the $\phi$ direction.

\section{Model Results}
\label{sec:model-fiducial}

\begin{figure*}
    \centering
    \includegraphics[width=1.\textwidth]{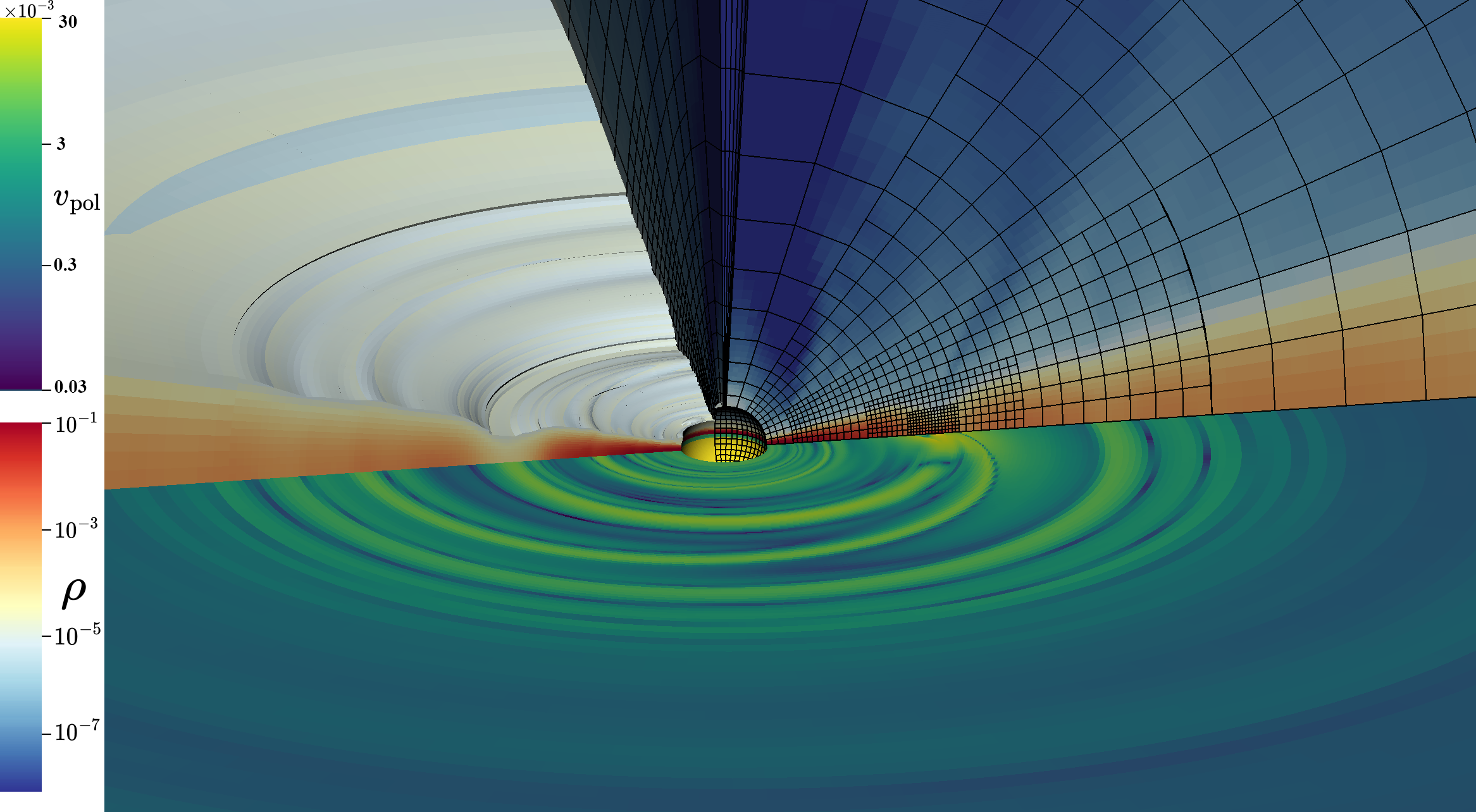}
    \caption{The simulation at the representative time t=100~\tp. The colormap in the top right part of the figure is the gas density distribution on the meridional plane passing through the planet, where the grid is the most refined (see the black grid, with each cell representing $4\times4$ simulation cells). Shown in the top left is a 3D view of an iso-density surface, highlighting the radial variation of the disk thickness induced by the planet and non-ideal MHD effects. The 2D colormap in the foreground (the lower half) plots the vertical velocity component on the midplane, illustrating the substructures developed in the disk.}
    \label{fig:grid}
\end{figure*}

To give a first impression of the simulation, we show in the top right part of Figure~\ref{fig:grid} the density distribution on the meridional plane through the planet at a representative time $t=100$~\tp, with the statically refined grid structure superposed. The top left panel shows a 3D view of an iso-density surface, highlighting the radial variation of the disk thickness induced by the planet and non-ideal MHD effects. The lower part of the figure shows the distribution of the vertical component of the velocity on the midplane, highlighting the substructures formed in the disk.

\begin{figure*}
    \centering
    \includegraphics[width=1.0\textwidth]{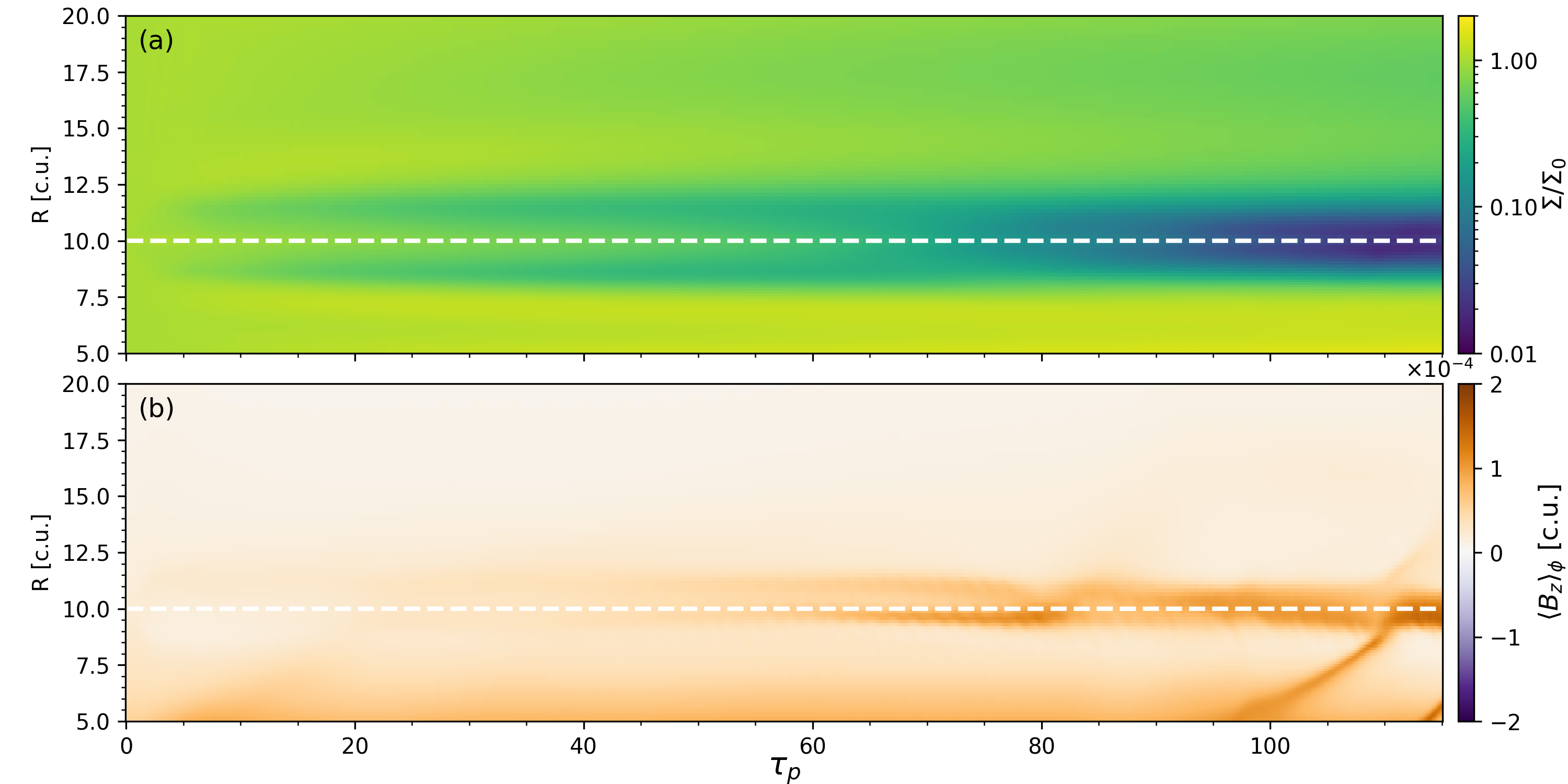}
    \caption{Space-time diagrams showing the evolution of (a) the gas surface density relative to its initial profile ($\Sigma/\Sigma_0$) and (b) the azimuthally averaged midplane vertical field strength ($B_z$). The location of the planet is marked with a white dashed line. }
    \label{fig:sigevo}
\end{figure*}

The surface density evolution over 115 planet's orbital periods (\tp) is shown in Figure~\ref{fig:sigevo}a, with the colormap illustrating the radial distribution of azimuthally averaged surface density normalized by the initial distribution $\Sigma(r)/\Sigma_0(r)$. The surface density at the gap center reaches about 1\% of the initial value after 100 \tp. Similar to \citet{2023ApJ...946....5A}, there's no apparent radial width asymmetry between the parts of the gap inside and outside the planet's orbit. This is also seen in the $Mj-\beta_4$ case from \citet{2023A&A...677A..70W}, which has the same planet mass and relative magnetic field strength as our setup. 
We also find the magnetic field is significantly strengthened in the gap (see Figure~\ref{fig:sigevo}b and also panel (d) of Figure~\ref{fig:fidp6ring} in Section~\ref{sec:gap}). The magnetic flux concentration is relatively stable at later times, with the concentrated flux staying near the gap center, similar to the $Mt1Am3$ case in \citet{2023ApJ...946....5A}. This behavior is different from the $Mj-\beta_4$ case of \citet{2023A&A...677A..70W}, where the location of magnetic flux concentration oscillated in the gap, leaving some isolated hot spot on their spacetime diagram, or their $Mj-\beta_3$ case, where the concentration diverges to the gap edges when the gap widens and becomes radially asymmetric.

\begin{figure*}
    \centering
    \includegraphics[width=1.0\textwidth]{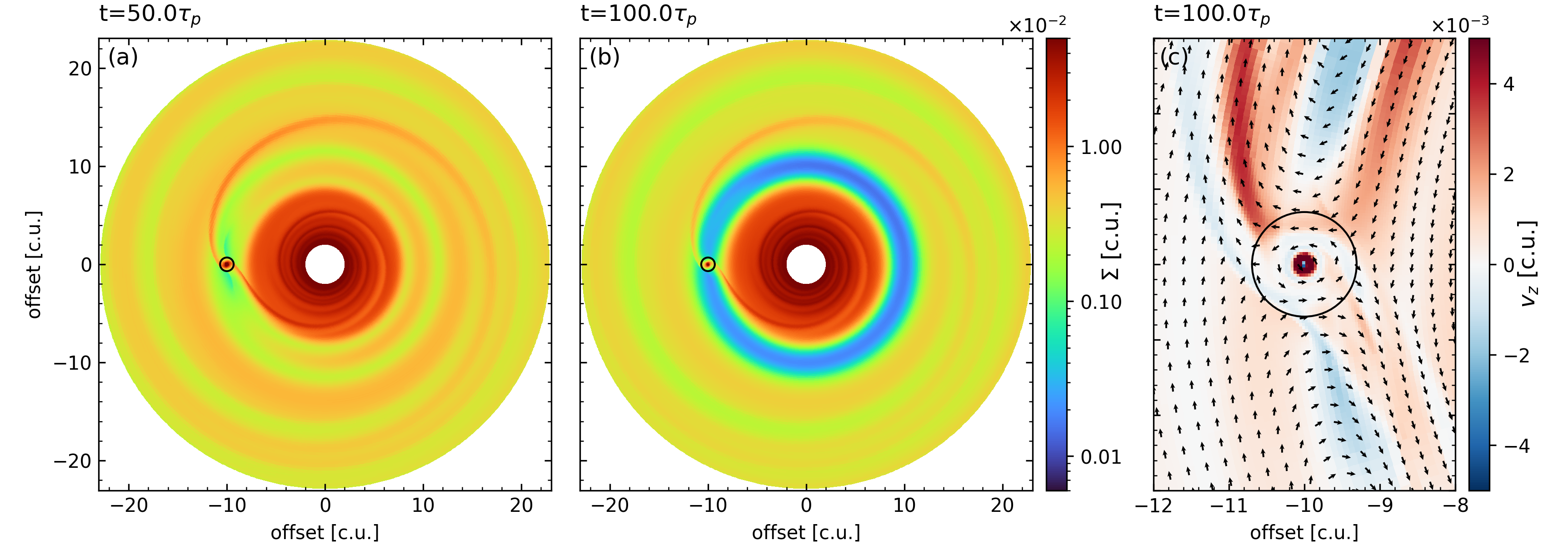}
    \caption{Snapshots at t=50 and 100~\tp. The color maps in panels (a) and (b) are the gas surface density in code units (c.u.). The black circle highlights the location and size of the planet's Hill sphere. Panel (c) shows a zoom-in view of the gas velocity structure, with the arrows representing the midplane velocity vectors in the planet's rest frame. The color map is the midplane vertical velocity.}
    \label{fig:sig3p}
\end{figure*}

We plotted the gas surface density snapshots at two representative times in Figure~\ref{fig:sig3p}. Panel (a) shows that the gap starts to be opened at t=50~\tp, with two prominent spiral density wakes, just as in a viscous disk. Both spiral wakes leave some secondary density perturbations, and the outer one seems to be more significant. There is a slight density bump in the gap since the gap is still getting deeper. The minimum density is seen at the leading and trailing edge of the planet. The gap has not shown azimuthal asymmetry at this point. In panel (b), the gap is well formed at t=100~\tp ~and in a steady phase, with a width roughly twice that of the planet's Hill sphere, highlighted with a solid circle. The density bump in the gap is gone, and only a few \% of gas remains at the gap center. A slight azimuthal asymmetry exists, with some mass concentration between L5 and the planet. The density perturbations ($R\sim 17$~c.u.) from the outer spiral wake are persistent, and almost form a full $2\pi$ gap.
Panel (c) of Figure~\ref{fig:sig3p} illustrates the velocity field on the midplane around the planet, with the arrows showing the classic horseshoe orbits and the rotating flow pattern inside the Hill sphere surrounding the dense circumplanetary disk and the colormap showing the midplane variation of the vertical velocity component. 

\begin{figure}
    \centering
    \includegraphics[width=0.45\textwidth]{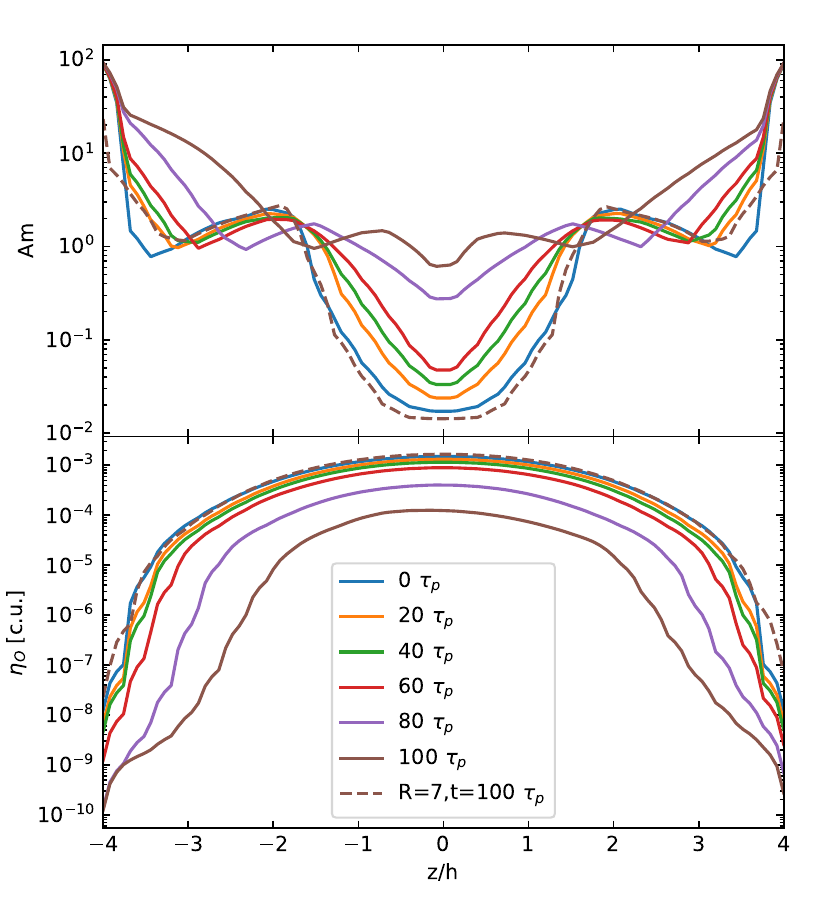}
    \caption{Evolution of the vertical profiles of the magnetic diffusivities at the gap center, up to 100\tp. The ambipolar diffusion Elsasser number $Am$ is in the upper panel, and the Ohmic resistivity $\eta_O$ is in the lower. Plotted are the vertical distributions of the azimuthally averaged diffusivities, excluding the region near the planet. The same color code is used in both panels to indicate the time of the snapshot. The dashed lines in both panels show the diffusion profiles of a region (R=7 in code unit, c.u. hereafter) that is not in the gap. }
    \label{fig:etavert}
\end{figure}

Figure~\ref{fig:etavert} shows the azimuthal average of the ambipolar diffusion Elsasser number $Am$ and the Ohmic resistivity $\eta_O$ obtained from the look-up table on a cylinder at the gap center's radius. Note the azimuthal average excludes the region within about 4 times the Hill sphere of the planet. Initially, the disk is most magnetically diffusive near the midplane (with the smallest Elsasser number), where the ionization level is low because of excessive recombination on the grain surface.  The reduction of magnetic diffusion is relatively modest in the first 60\tp, with a factor-of-3 increase in $Am$ and less than $50\%$ decrease in $\eta_O$. In contrast, by  t=100\tp, the $Am$ near the midplane has increased by nearly two orders of magnitude, yielding an almost constant vertical profile with values close to unity, which, coincidentally, is similar to the fixed $Am$ value adopted by both \citet{2023ApJ...946....5A} and \citet{2023A&A...677A..70W} in the disk. The Ohmic resistivity profile retains its initial shape, with a strength reduced by an order of magnitude. We also overplotted the diffusion profiles at a radius not in the gap, where the Elsasser number and Ohmic resistivity retain their initial distributions (dashed line) at 100\tp. Unlike the two previous studies, the jump in magnetic diffusivities from the rest of the disk to the gap in our setup implies some interesting gas dynamics at the gap edge (see Section~\ref{sec:gap}). Figure~\ref{fig:etavert} demonstrates that mass depletion makes the gap region better magnetically coupled compared to the initial disk midplane, which has implications on its dynamics. 

\begin{figure}
    \centering\includegraphics[width=0.45\textwidth]{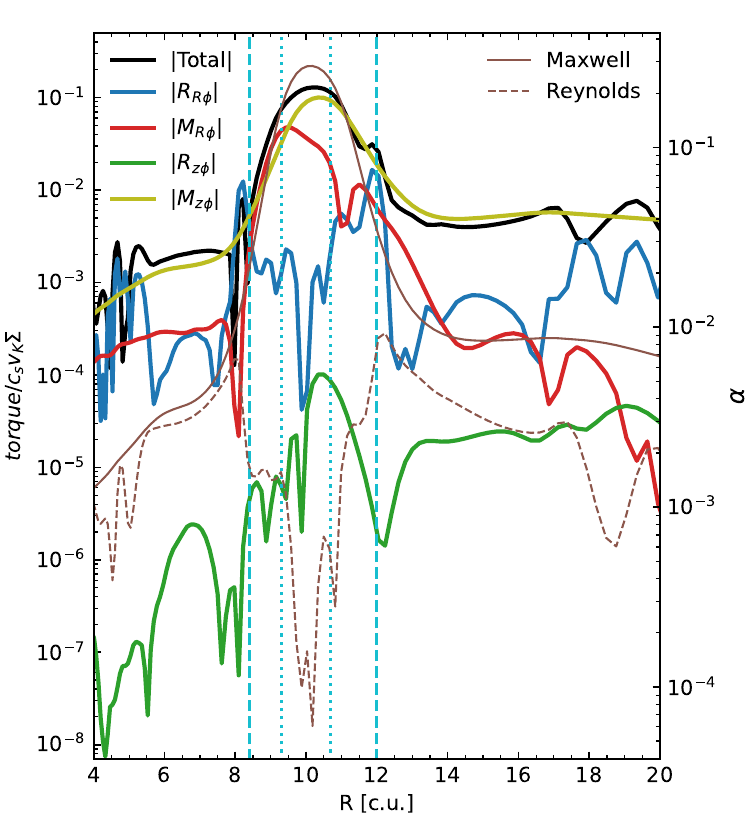}
    \caption{Radial profile of the absolute values of the four torques $\mathcal{R}_{R\phi}$ (blue), $\mathcal{M}_{R\phi}$ (red), $\mathcal{R}_{z\phi}$ (green), $\mathcal{M}_{z\phi}$ (yellow), and the total torque (black). All torques are normalized by $c_sv_K\Sigma$ and azimuthally-averaged and temporally-averaged from t=90\tp~ to 100\tp. Scaled by the right y-axis, the thin brown solid and the thin brown dashed lines are the Maxwell $\alpha$ and the Reynolds $\alpha$, respectively. The cyan vertical lines mark the area of the corotation region (dashed) and the planet's Hill sphere (dotted).}
    \label{fig:stress}
\end{figure}

We choose to perform diagnostics of the simulation in cylindrical coordinates to simplify angular momentum analysis. The azimuthally averaged $R\phi$ and $z\phi$ stresses govern the disk angular momentum transport:
\begin{eqnarray}
    {T}_{R\phi}=\frac{1}{2\pi}\int_0^{2\pi}\left[\rho v_R \left(v_\phi -\bar{v}_\phi\right)-B_RB_\phi\right] \d\phi\\ 
    {T}_{z\phi}=\frac{1}{2\pi}\int_0^{2\pi}\left[\rho v_z \left(v_\phi-\bar{v}_\phi\right)-B_zB_\phi\right] \d\phi
\end{eqnarray}
In both formulas, $\bar{v}_\phi$ is the azimuthally averaged toroidal velocity, the first term inside the integral denotes the contribution from Reynolds stress, and the second term represents Maxwell stress. In a (quasi-)steady accretion disk, the accretion is driven
by the radial gradient of $T_{R\phi}$, the
difference of the $T_{z\phi}$ between the top (marked as $t$ as integration limit, where $\theta-\pi/2=-0.25$) and
bottom disk surfaces (marked as $b$ as integration limit, where $\theta-\pi/2=0.25$), and the torque from the planet $\Gamma_p=\frac{1}{2\pi}[\int_b^t\int_0^{2\pi}\rho (\bm{r}_p\times\bm{\nabla}\Phi_p)\d\phi \d z]_z$, namely,
\begin{eqnarray}
  \dfrac{\dot{M}_\acc \Omega_\K}{4\pi}
  &\simeq&\frac{1}{R}\dfrac{\partial}{\partial R}
  \left( R^2\int_{\rm b}^{\rm t} T_{R\phi}\ \d z \right)
  +R \left[ T_{z\phi}\right]_{\rm b}^{\rm t}\ - \Gamma_p\nonumber\\
  &=&\mathcal{R}_{R\phi}+\mathcal{M}_{R\phi}+\mathcal{R}_{z\phi}+\mathcal{M}_{z\phi}-\Gamma_p
\label{eq:stress}
\end{eqnarray}
Here $\dot{M}_\acc$ is the accretion rate within the disk, and $v_K$ is Keplerian orbital velocity. We used $\simeq$ as we approximate the disk's $v_\phi$ by the Keplerian speed, since the azimuthally averaged $v_\phi$ is within $4\%$ of $v_K$ at the midplane. Then we have $\dot{M}_\acc \partial (r\bar{v}_\phi)/\partial r\simeq\dot{M}_\acc \partial (rv_K)/\partial r = \dot{M}_\acc v_K/2$ for the radial transport of angular momentum under a constant mass accretion rate. t is worth noting that this approximation introduces a deviation of up to 50\% at the gap edge due to the alteration of the rotation curve caused by the planet. However, this does not affect our analysis, as we will only compare the terms on the right-hand side of Equation~\ref{eq:stress}.
Following \citet{2023A&A...677A..70W}, we divided the stress terms on the right-hand side of Eq.~\ref{eq:stress} into its Reynolds and Maxwell component, written in the second row. The four terms that can contribute to the total planet-free torque are the radial Reynolds ($\mathcal{R}_{R\phi}$) and Maxwell ($\mathcal{M}_{R\phi}$) torques and the $z\phi$ Reynolds ($\mathcal{R}_{z\phi}$) and Maxwell ($\mathcal{M}_{z\phi}$) torques. Figure~\ref{fig:stress} shows the contribution of the absolute value of these four torques to the disk accretion as a function of the distance to the star. All torques are scaled to $c_sv_K\Sigma$ to get the dimensionless $\upsilon$ parameter defined in \citet{2021A&A...650A..35L}. The Maxwell wind torque ($\mathcal{M}_{z\phi}$, yellow line) is the dominant term in the whole radial range, which means the angular momentum transport is mainly through magnetic wind loss through the disk surface. The next two contributions are both torques due to radial stresses, the Maxwell ($\mathcal{M}_{R\phi}$, red line) and Reynolds ($\mathcal{R}_{R\phi}$, blue line) components, which are comparable outside the gap. With the gap edges marked by two vertical dashed lines, we find that both Maxwell torques are enhanced by about an order of magnitude inside the gap, while the radial Reynolds torque is suppressed relative to its Maxwell counterpart. The magnetic flux concentration and gas depletion in the gap are both contributing to this difference. The Reynolds wind torque ($\mathcal{R}_{z\phi}$, green line) contributes little to the total accretion.

The relative strength of $R\phi$ stress can be characterized by the equivalent
Shakura-Sunyaev $\alpha$ parameter,
defined as the ratio of the vertically integrated $R\phi$ stress to the thermal pressure in the disk ($\pm3.5h$):
\begin{eqnarray}
 \alpha_{Max}=\left|\left.\int_{\rm b}^{\rm t}\int_0^{2\pi}{B_RB_\phi}\d\phi\d z \right| \middle/\int_{\rm b}^{\rm t}\int_0^{2\pi}P_{\rm gas}\d\phi\d z\right.  \\
 \alpha_{Rey}=\left|\left.\int_{\rm b}^{\rm t}\int_0^{2\pi}{\rho v_R(v_\phi-\bar{v}_\phi)}\d\phi\d z \right|  \middle/\int_{\rm b}^{\rm t}\int_0^{2\pi}P_{\rm gas}\d\phi\d z\right.
\end{eqnarray}
The Maxwell and Reynolds $\alpha$'s are plotted in Figure~\ref{fig:stress}, with scaling given on the right y-axis. The magnitude of $\alpha_{Rey}$ is just a fraction of $\alpha_{Max}$, and they show a similar trend inside the gap as $\mathcal{R}_{R\phi}$ and $\mathcal{M}_{R\phi}$. The planet creates two large-scale structures: the gap and two spiral-density waves. Because $\mathcal{R}_{R\phi}$ relies on the radial gradient of the $R\phi$ stress, those two structures can boost the contribution from the Reynolds component. The sharp drop of $\alpha_{Rey}$ at the gap edges results in the highest value of $\mathcal{R}_{R\phi}$, almost dominating the total torque at these locations.

\begin{figure}
    \centering
    \includegraphics[width=0.45\textwidth]{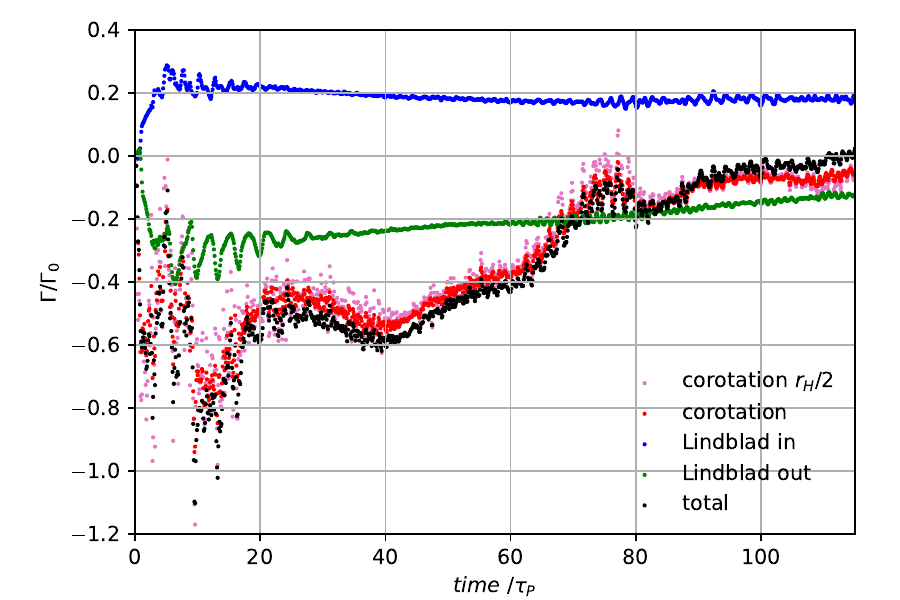}
    \caption{Evolution of the migrational torques acting on the planet in the fiducial setup, normalized by $\Gamma_0$. Plotted are the corotational torque (red dots), Lindblad torques from the disk gas outside the outer edge (green) and inside the inner edge (blue) of the corotation region, and the total torque (black). The pink dots represent the corotational toque calculated with half tapering radius.}
    \label{fig:fid_torq}
\end{figure}

The Type II migrational torque is divided into the corotational and Lindblad torque. We first identify the corotational region by checking the velocity vectors in the midplane, and the region is set by $8.4<R<11.8$, roughly five times the planet's Hill radii. The Lindblad torque $\Gamma_L$ is simply the sum of the torque of all cells outside this cylindrical zone, and we calculate the torque from the inner and outer disk separately. For the torque from the corotational region $\Gamma_C$, in addition to the same softening length used in Eq.\ref{eq:potential}, we minimize the effect of the gas inside the Hill sphere by multiplying a taper function $f_H=1-e^{-|\bm{r}-\bm{r_p}|^2/r_t^2}$, where $r_t$ is the tapering radius. We compare two choices of $r_t$: the fiducial $r_t = r_H$ used in \texttt{FARGO} \citep{2000A&AS..141..165M} and $0.5r_H$. While the latter exhibits higher fluctuations, the overall behavior of both tracks remains consistent and overlaps. $\Gamma_C$ dominates the total torque when $t<75$\tp, as shown 
in Figure~\ref{fig:fid_torq}. We normalized the torque by the characteristic scaling~\citep[e.g.,][]{2010MNRAS.401.1950P} $\Gamma_0=q^2(h/r)^{-2}R_p^4\Omega_p^2\Sigma_p$. Both $\Gamma_C$ and $\Gamma_L$ are negative torques in this early phase. After 100\tp, the corotational torque drops to near zero, and the Lindblad torque is reversed to be slightly positive, and so is the combined total torque.

\begin{figure}
    \centering
    \includegraphics[width=0.45\textwidth]{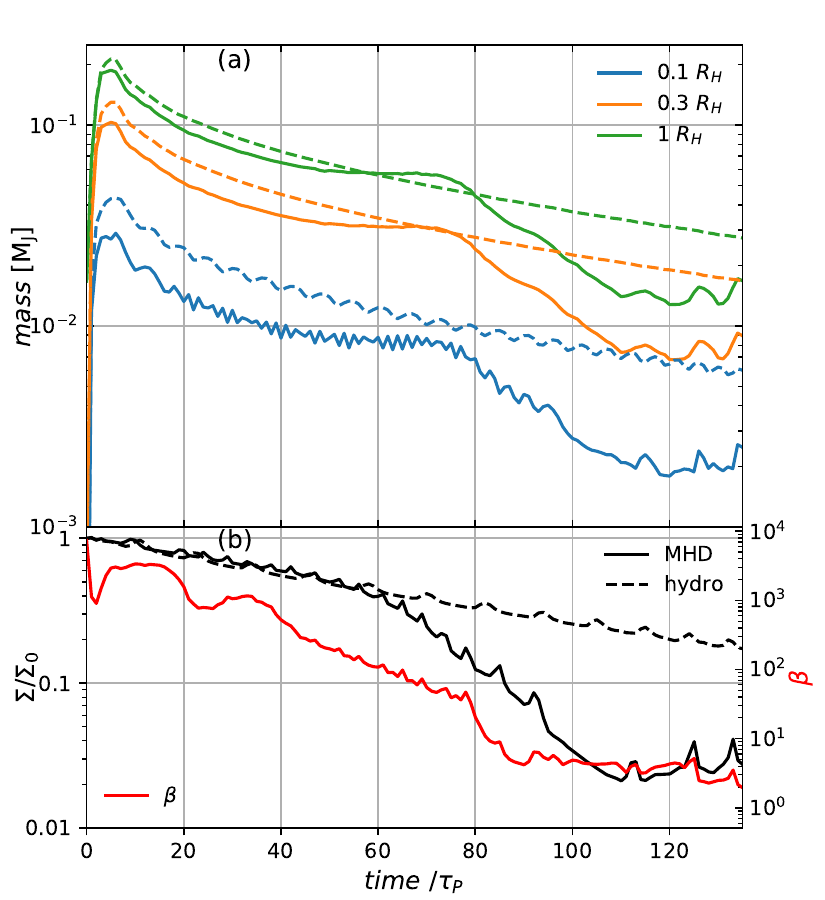}
    \caption{Panel (a): Time evolution of gas mass enclosed within spheres of 1, 0.3, and 0.1 $R_H$ (Hill radius) centered on the planet. The solid lines represent the fiducial MHD case, and the dotted lines are from the hydro case with viscous $\alpha=0.001$. Panel (b): The black lines are the relative gap depths for the MHD (solid) and hydro (dashed) cases, and the red line is the vertically averaged plasma $\beta$ at the gap center in the MHD case, with the scaling shown on the right y-axis. }
    \label{fig:hillmass}
\end{figure}

An intriguing effect of the MHD wind is that it can suppress the planet accretion by removing gas in the mass reservoir. Though we don't have a mass sink for proper long-term planetary accretion study, we can monitor the gas concentration near the planet. We do so by computing the masses enclosed within three spheres centered on the planet, with radii of 1~$R_H$, 0.3~$R_H$, and 0.1~$R_H$. We take mass within 0.3~$R_H$ as a proxy for the circumplanetary disk (CPD) mass (based on an inspection for the surface density distribution near the planet; see, e.g., Figure~\ref{fig:sig3p}b), shown as orange lines in Panel (a) of Figure~\ref{fig:hillmass}. The same mass measurement is performed for the hydro case with viscous $\alpha=0.001$, plotted as dashed lines of the same color. We use black lines in Panel (b) to represent the gap depth $\Sigma/\Sigma_0$ at $R=R_p$. We did not exclude the gas in the planet's vicinity when measuring $\Sigma$. This may cause some spikes in the time evolution track, but not the general trend. The mass time evolution follows a smooth curve in the viscous disk. In the MHD case, a significant increase in the rate of mass loss within the Hill sphere is observed compared to the hydro case after $\approx75$ $\tau_p$. We also observed the drop of plasma $\beta$ (vertically averaged from -h to +h in the midplane) at the same epoch, plotted as the solid red line in Panel (b), with the scaling shown on the right y-axis. The gap depth time evolution has a similar trend, i.e., the MHD case closely follows the hydro case in the first $\approx60$ orbits. The MHD gap opening process in this early stage behaves similarly to a pure hydro process without a significant magnetic field concentration. After the magnetically driven accretion starts in earnest, it quickly depletes the gap region with a high radial velocity. The result is a less massive CPD compared to the hydro case. We cannot confirm if this is due to direct wind loss from the CPD itself. Nevertheless, the strong magnetization and better field-matter coupling in the gap region and the associated fast accretion have a negative impact on the mass reservoir for the planet growth.

\subsection{Gas Dynamics of the Gap}
\label{sec:gap}

\begin{figure*}
    \centering
    \includegraphics[width=1.0\textwidth]{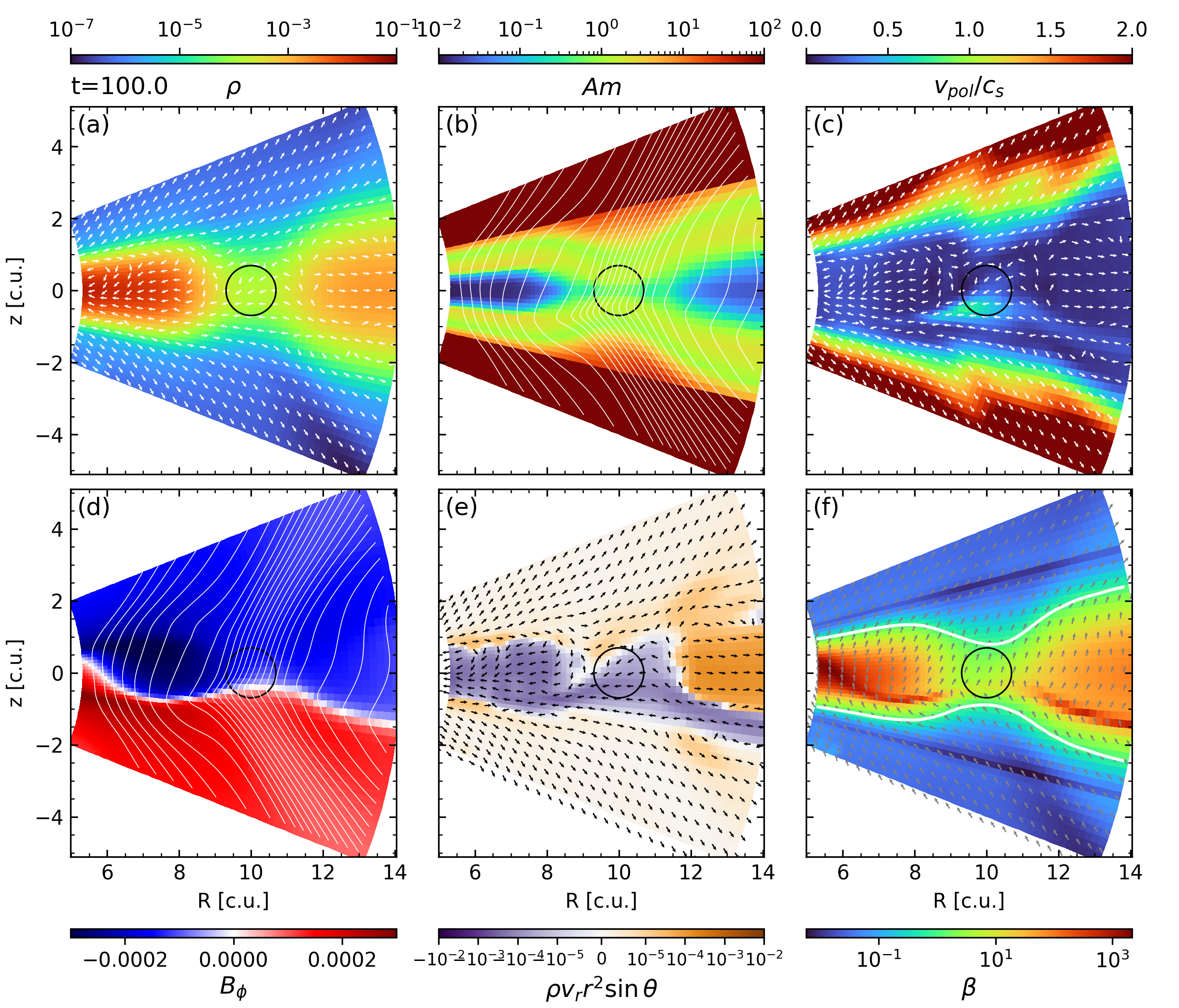}
    \caption{Azimuthally averaged gas properties at the gap excluding the planet. Panel (a): gas density (in code units) with white arrows representing velocity vectors in the poloidal plane; (b): ambipolar Elsasser number and poloidal magnetic fields (solid white lines); (c): poloidal velocity divided by the local sound speed with white arrows representing velocity vectors in the poloidal plane. Panel (d): Azimuthal magnetic field $B_\phi$ and poloidal magnetic fields (solid white lines);(e): effective radial mass flux per unit area and poloidal velocity vectors (black arrows); (f): plasma $\beta$ and magnetic field vectors in grey arrows, with the white contour marking $\beta=1$. The solid black circle in each panel shows the planet's Hill sphere. See the supplementary online material for an animated version of this figure.}
    \label{fig:fidp6ring}
\end{figure*}

To better illustrate the global structure of the gap, we azimuthally averaged the gas properties excluding the region close to the planet (15$^\circ$ sector on each side of the planet). 
Figure~\ref{fig:fidp6ring} (a) and(b) show that the distributions of the gas density and ambipolar Elsasser number $Am$ are broadly mirror-symmetric with respect to the disk midplane. It is not the case for the poloidal magnetic and velocity fields, however. The $Am$ profile roughly follows the gas density. While $Am$ generally anti-correlates with $\rho$, there is an $Am$ ``valley" just below the disk surface (see the greenish layer below the reddish surface layer in Figure~\ref{fig:fidp6ring}b), which resembles the power-law prescription proposed by \citet{2018MNRAS.477.1239S} where the $Am$ inside the disk decreases with a decreasing density as a result of balancing the recombination of the dominant ions and electrons from cosmic ray ionization. In panel (c), there's a region of high poloidal velocity ($>0.5c_s$) just below the midplane. Its spatial distribution is associated with the kinks on poloidal magnetic field lines in panel (b). This is also reflected in panels (d) and (f), where the fast accretion happens in the layer where $B_\phi$ changes sign and the plasma $\beta=2P_{gas}/B^2$ has a local maximum (because of a vanishing $B_\phi$). This layer lies significantly below the midplane, indicating that the accretion is not mirror-symmetric with respect to the midplane (see panel [e]), unlike the density distribution. Note again that the gap is significantly magnetized, with a plasma-$\beta$ typically between 1 and 10, much less than the initial midplane value of $10^4$.

\begin{figure*}
    \centering
    \includegraphics[width=1.0\textwidth]{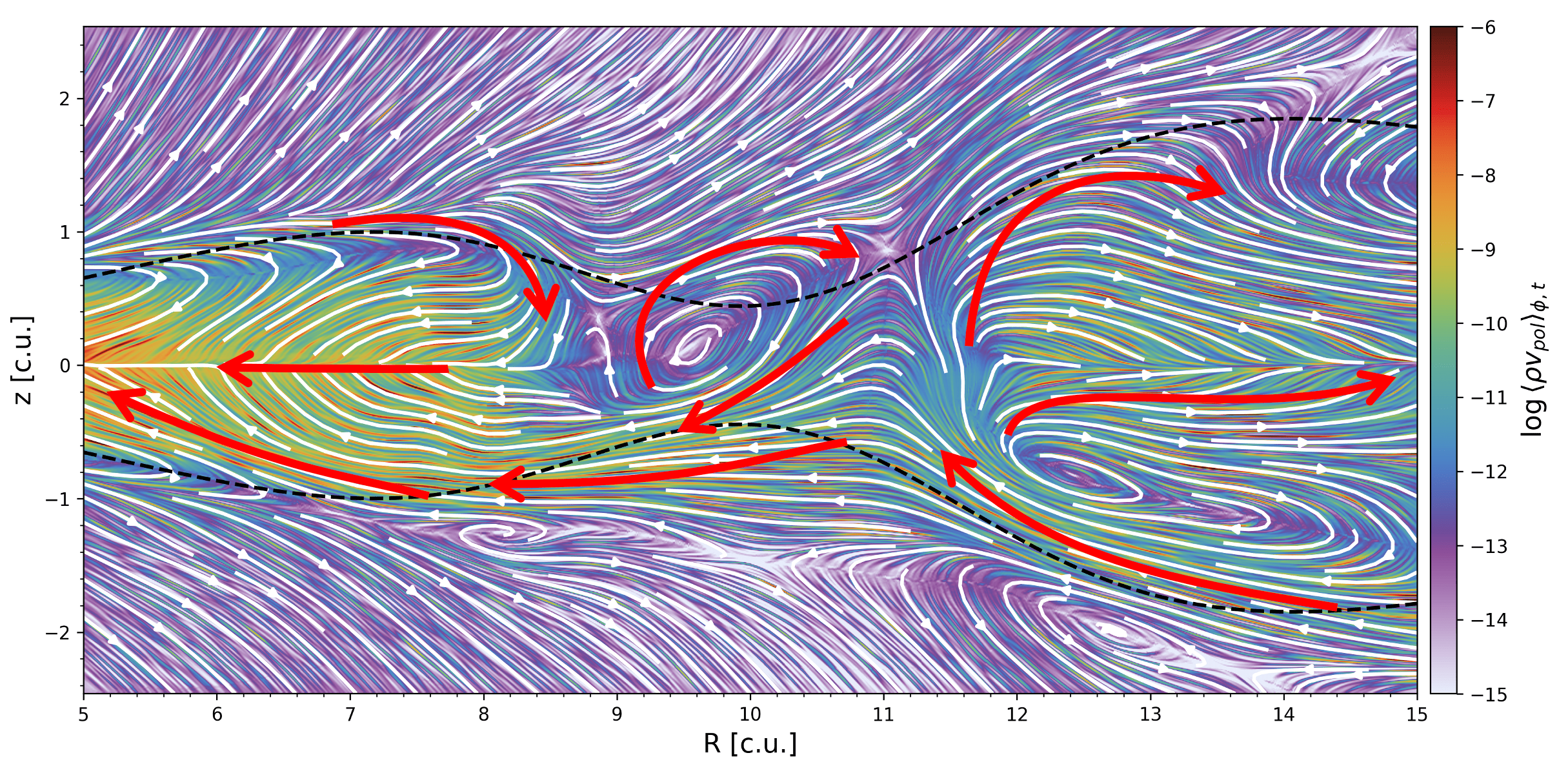}
    \caption{Meridional flows in and around the gap, averaged in azimuth (excluding the planet) and from t=90\tp~to t=100\tp. The flow structure is highlighted with LIC and white streamlines. The black solid lines are the ``emission surface'' with $z_0=0.2$ to be used in Section~\ref{sec:observe}. The red arrows assemble a schematic view of the overall flow structure. }
    \label{fig:fidstream900}
\end{figure*}

In Figure~\ref{fig:fidstream900}, we present the intricate meridional flow structure within the gap, visualized through line integral convolution (LIC). Beyond the noticeable outflows at the disk's upper and lower surfaces, a particularly prominent feature is the persistent one-sided accretion stream. This stream originates from the lower surface beyond the outer gap edge ($>14$~c.u.), traverses almost uninterrupted across the bottom half of the gap, and eventually converges with the accretion flow within the inner gap edge ($<6$~c.u.). Notably, several elongated vortices form between the accretion layer and the disk wind emanating from the lower surface. Within the gap, near the disk's midplane, a significant circulation pattern is observed above the accretion layer. This pattern denotes a transition from accretion below to outflow above. Additionally, the wind is launched from a lower surface in the gap (see the poloidal velocity in Figure~\ref{fig:fidp6ring}c), especially for the upper side. This allows the disk wind to directly impinge on the ridge of the outer gap edge, which could intensify decretion at this location in conjunction with planetary torque. The outgoing flow at the outer edge is vertically extended across the disk midplane, creating another elongated vortex between itself and the accretion stream near the bottom disk surface.

\begin{figure}
    \centering
    \includegraphics[width=0.45\textwidth]{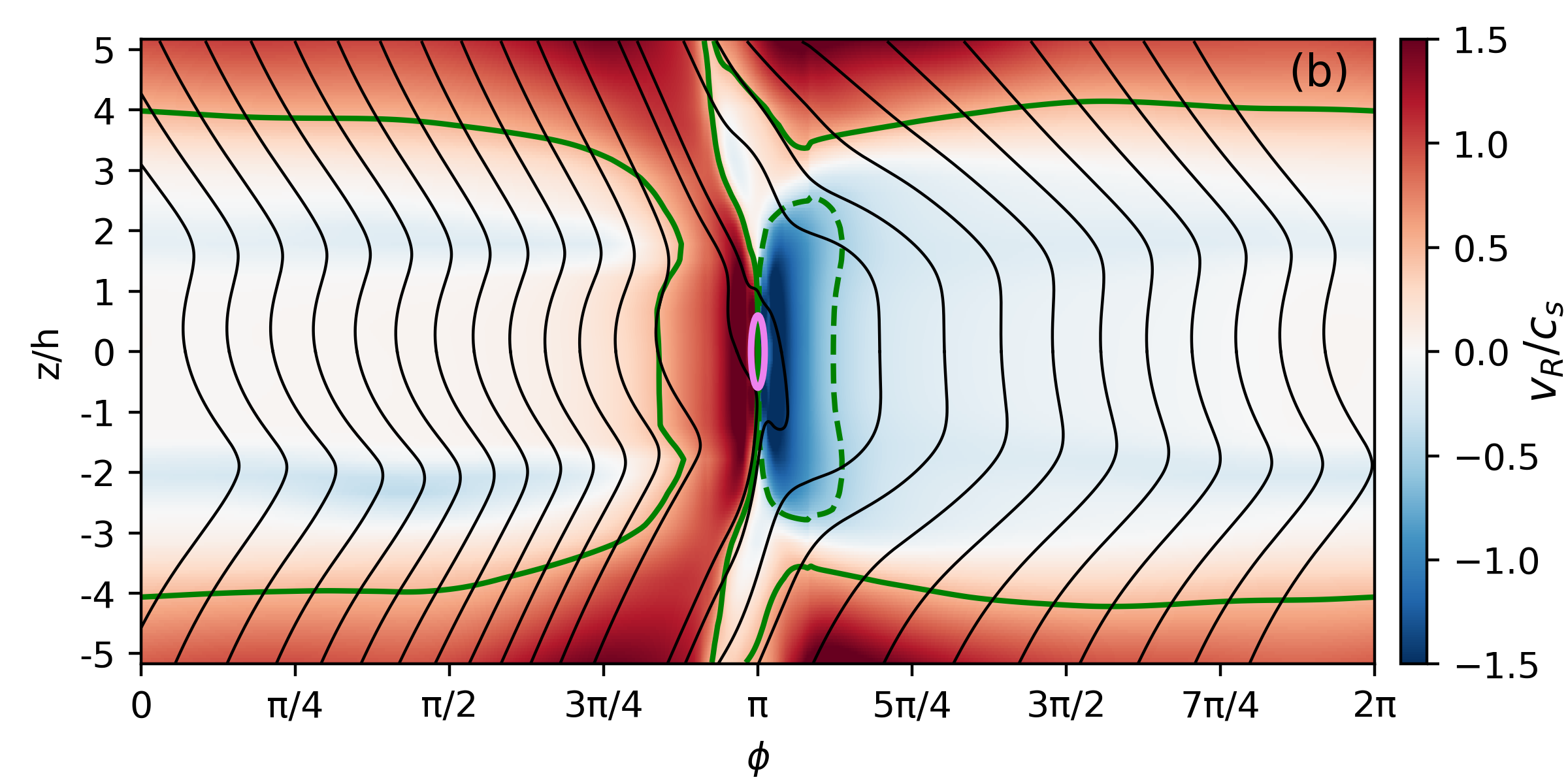}
    \includegraphics[width=0.45\textwidth]{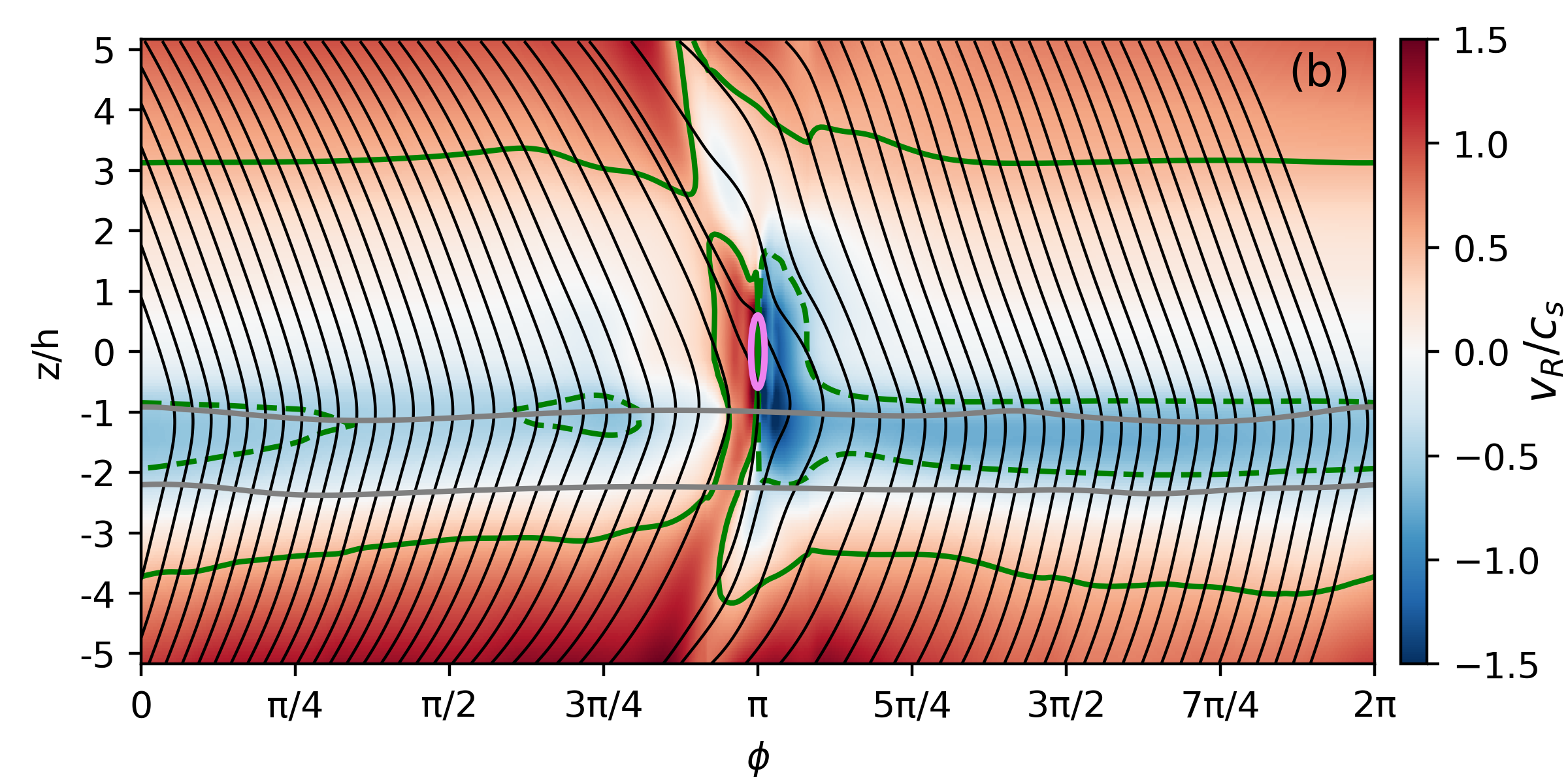}
    \caption{Gas radial velocity scaled to the midplane sound speed of a $z-\phi$ slice at the planet's orbit. Panel (a) is a time average from 40 to 50 $\tau_p$, and panel (b) is from 90 to 100 $\tau_p$. The magnetic field in the $z-\phi$ plane is plotted as black solid lines, and both panels used the same magnetic flux spacing between two adjacent lines, showing a stronger field at the later time frame. The green contours mark the location where $v_R = \pm 0.5~c_s$, with dashed lines representing negative values. The violet contour at the center indicates the planet's Hill sphere, which mainly highlights the location of the planet. It is not spherical because the height and the azimuthal angle axes are not plotted to the same scale. The area between the two roughly horizontal lines in (b) has the highest radial current density $j_r$.}
    \label{fig:azcutvrcs}
\end{figure}

To determine if the accretion layer spans the entire azimuth of the gap and to directly compare the accretion structure before and after the gap's full formation, we analyzed the radial velocity distribution on a cylindrical surface parallel to the disk's rotation axis at R=10. For comparison, Figure~\ref{fig:azcutvrcs}(a) presents the time-averaged data from 40 to 50~\tp. Beyond the planet's immediate vicinity, accretion concentrates at $\pm2h$, just below the disk surfaces, aligning with the 2D findings of \citet{2019ApJ...874...90W}. This accretion layer spatially correlates with the locations of significant field lines bending, which in turn are linked to significant wind stress, $M_{z\phi}$. On the midplane (where $z=0$), the azimuthal distribution of $v_R$ changes sign at the planet's position, with a lower absolute value at locations away from the planet than in the accretion layer. Gas flows inward on the leading side  ($\phi>\pi$) and outward on the trailing side, characteristic of the horseshoe region in a viscous disk. Only the gas at the wind base (approximately $\pm4h$) and adjacent to the planet reaches a radial velocity exceeding $0.5~c_s$.

\begin{figure}
    \centering
    \includegraphics[width=0.45\textwidth]{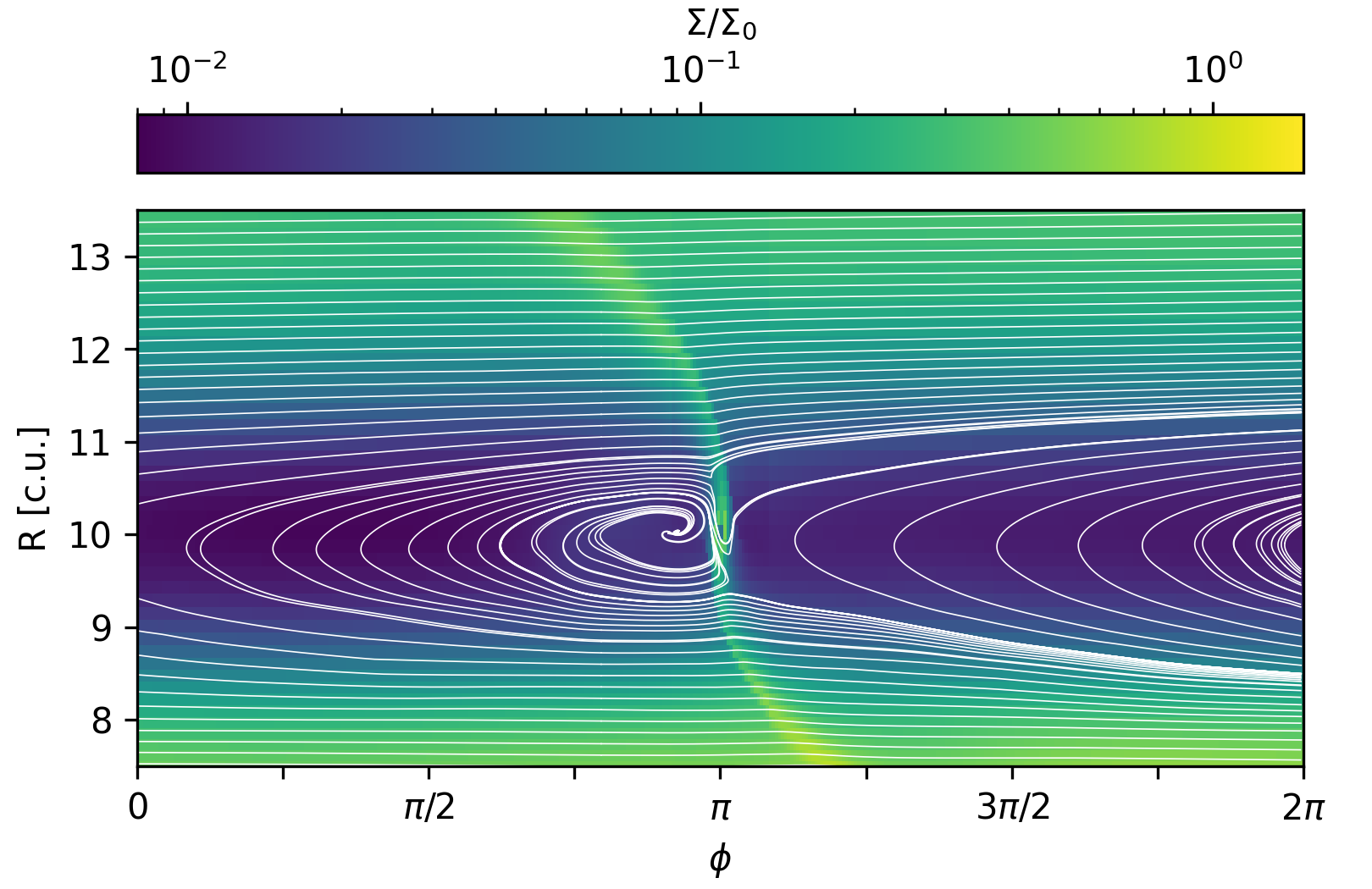}
    \caption{Streamlines around the gap at t=100\tp, plotted with the gas velocity at the layer with maximum $j_r$. The background colormap is the gas surface density variation $\Sigma/\Sigma_0$.}
    \label{fig:horseshoe}
\end{figure}

After the gap fully opens, the kinks in the magnetic field lines that are originally near both disk surfaces ($\pm2h$) merge around $-h$ (Figure~\ref{fig:azcutvrcs}b). The azimuthal component of the Lorentz force ${F}_{L,\phi}\propto (\bm{j}\times\bm{B})_\phi\approx j_r B_{\theta}$ (since $B_{\theta}\gg B_r$), where $j_r\propto (\nabla\times\bm{B})_r=\frac{1}{r\sin\theta}(\frac{\partial}{\partial\theta}(B_\phi\sin\theta)-\frac{\partial B_\theta}{\partial\phi})$. Since $B_\theta$ varies slowly in the disk along the $\theta$~direction, the strength of $j_r$ resembles the magnetic torque\footnote{We use spherical coordinate here instead of cylindrical coordinates for two reasons: first, the spatial derivation is more accurate in spherical form as it follows the grid structure; second, the strongest current deviates more than one scale height from the midplane, which makes the spherical $r$ better represents the direction of accretion than the cylindrical $R$. }. The gray contours in Figure~\ref{fig:azcutvrcs}(b) marks $j_r=5\times10^{-4}$ in code units, roughly 1/3 of the maximum value. Between the gray contours, the region with the highest $j_r$ is closely associated with the rapid accretion layer. After t=80~\tp,  the break from the mirror symmetry across the planet in the azimuthal distribution of the radial mass accretion becomes prevalent, significantly reducing the horseshoe region's size on the planet's trailing side. The streamlines in Figure~\ref{fig:horseshoe}, drawn based on the velocity of cells with the highest $j_r$, further confirm that the Lorentz force directly enables the fast accretion layer. A minor over-density is observed between the planet and the trailing Lagrangian point L5, likely due to material accumulation by the vortex from the reduced horseshoe orbit. The pronounced asymmetry in streamlines is also observed in case $Mj-\beta 3$ from \citet{2023A&A...677A..70W}, where the horseshoe region narrows to a zone with limited azimuthal extent, potentially leading to a similar dust distribution. \citet{2023MNRAS.523.2630W} proposes that these asymmetric dust clumps could serve as observational indicators of wind-driven accretion.

\subsection{Observational Implications}
\label{sec:observe}

\begin{figure}
    \centering
    \includegraphics[width=0.45\textwidth]{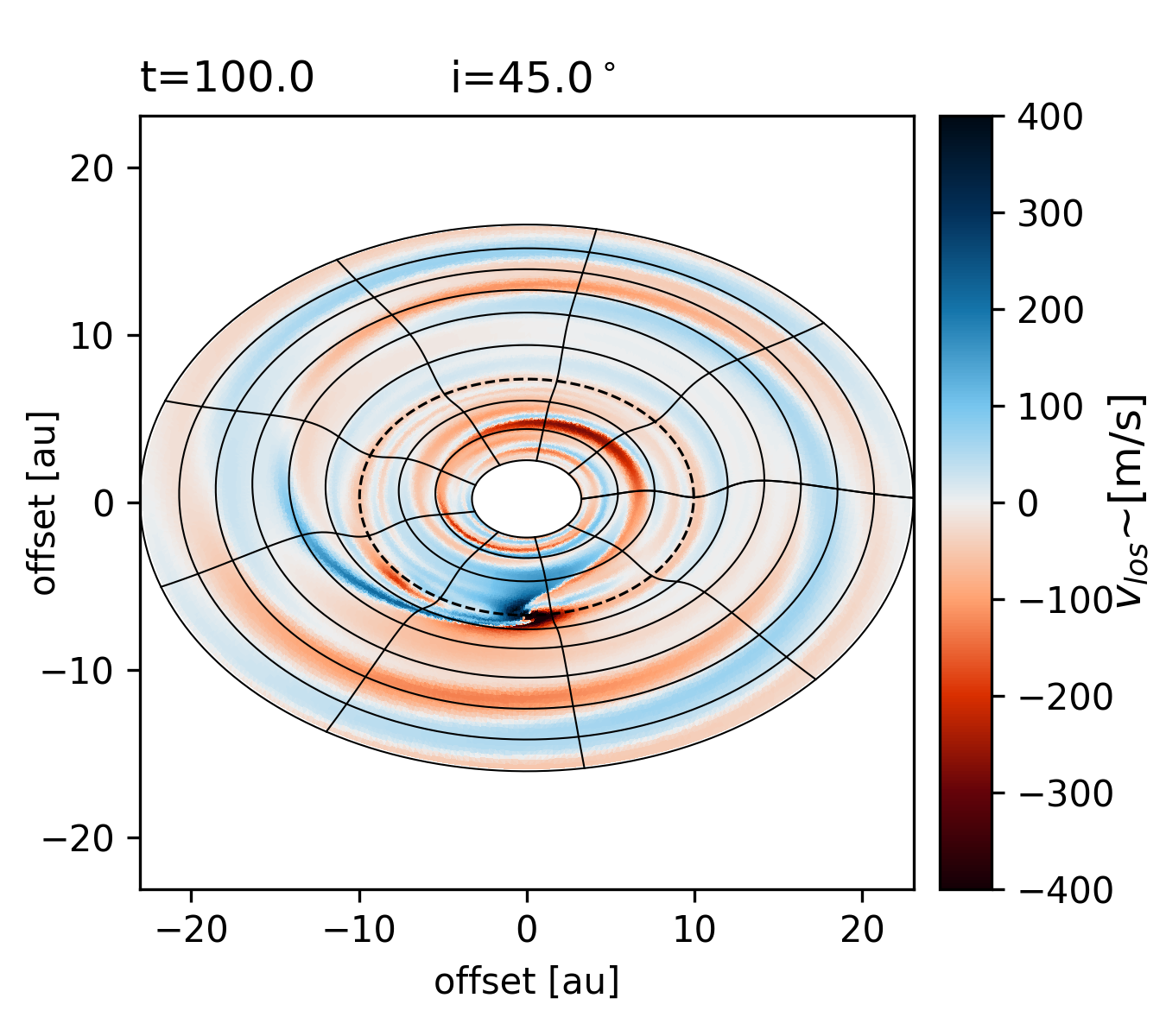}
    \caption{Line-of-sight gas kinematics extracted at the parametric emission surface, with an inclination of 45 degrees, after subtracting mean azimuthal velocity. The dashed circle marks the gap center. Note this is not the residual velocity after subtracting the three axisymmetric velocity components.}
    \label{fig:vlos}
\end{figure}

In this section, we extract the gas kinematics in a way similar to that used for recent ALMA line observations \citep[e.g.][]{2021ApJS..257...18T,2021A&A...650A.179I}. Unlike \citet{2023MNRAS.523.4883H}, who marked the emission surface with a constant column density from the top of the simulation domain, we choose to model the shape of the emission surface using the analytic function suggested in the molecular line data analysis tool \texttt{eddy}~\citep{eddy}:
\begin{equation}
    z\left(R\right)=z_0[1-g(R)](R-R_c)^\Phi\exp[(\frac{R-R_t}{R_t})^{q_t}]
    \label{eq:surf}
\end{equation}
We choose $z_0=\pm0.1,0.2,0.3$, $R_c=1$ c.u., $\Phi=1$, $q_t=2$, and $R_t=9$ c.u. for our model. Within the tapering radius $R_t$, $z_0=\pm0.1,0.2,0.3$ each translates to 1.7, 3.5, and 5 scale heights above/below the midplane. The vertical modulation $g(R)=d_0\exp[-(R-R_p)^2/(2w_h^2)]$ is a Gaussian function that adds the gap shape to the smooth surface, where $d_0=0.75$ is the maximum relative gap depth in $z/R$. $R_p=10.$ c.u. is the planet's orbital radius, and $w_h=1.6$ is the half-width. The drop in emission surface at the gap is found in disks like AS209 \citep{2023ApJ...950..147G}, HD163296, and MWC480, after carefully masking the channel emission \citep{2023A&A...669A.126P,2023A&A...674A.113I}. We choose the median value $z_0=\pm0.2$, i.e., $z\sim\pm3.5h$ where $R<R_t$ as the simulated emission surface (see the dashed lines in Figure~\ref{fig:fidstream900}). The velocities $v_r$, $v_\theta$, and $v_\phi$ are averaged over one scale height, weighted by the gas density. We project the velocity vectors to the line of sight (LOS) at an inclination $i=45$ deg,  with the planet lying on the vertical line on the near side of the inclined disk. Assuming one code length unit equals one au, we get the 2D LOS velocity distribution in $m/s$ in Figure~\ref{fig:vlos}, with meshes highlighting the shape of the ``emission surface.'' Note that this parametric surface does not accurately represent the real-life emission surface of any specific molecule. Instead, it serves as a demonstration of the velocity distribution on a layer distorted by a deep gap. Factors such as chemical abundance, gas temperature, and velocity gradient all contribute to determining the shape and location of the emission surface. While the MHD disk wind could significantly alter the gas column density, its impact on the gap emission surface is expected to be less pronounced due to the substantial difference in the line-of-sight velocities between a rotating disk and a radial outflow-dominated wind (i.e., the wind does not contribute much to the optical depth and hence the location of the $\tau=1$ emission surface at the disk's line-of-sight velocities).
Note this follows the same scaling from the look-up table of magnetic diffusivities, in which the planet is also placed at 10 au. In Figure~\ref{fig:azcutvrcs}, the $v_R$ sign transition near the planet can extend up to two scale heights. It shows up clearly in the kinematics of the emission surface near the planet as a sharp transition from redshift on the leading side (to the right) of the planet to blueshift on the trailing side. 
Moving away from the planet, the main source of non-axisymmetric features comes from the spiral density waves. They generate notable variations in the LOS velocity map. 

\begin{figure}
    \centering
    \includegraphics[width=0.45\textwidth]{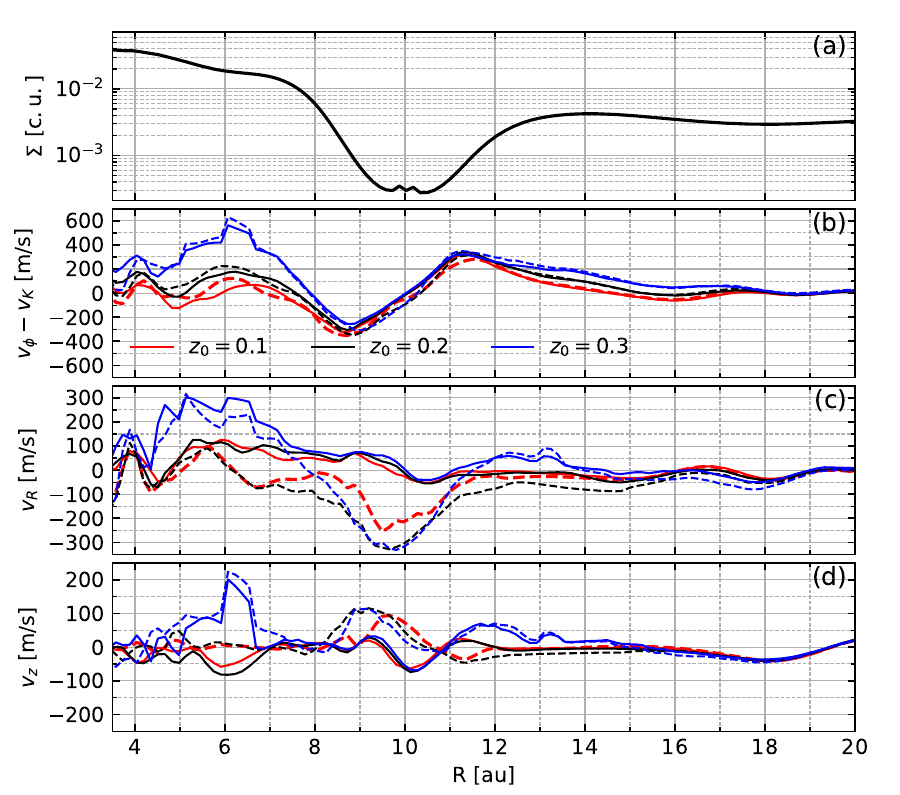}
    \caption{1D (axisymmetric) velocity profile for three parametric emission layers: $z_0=0.1$ (red), $z_0=0.2$ (black), and $z_0=0.3$ (blue) at t=100\tp. The top panel (a) is the gas surface density for reference, and panels (b), (c), and (d) are $\phi$ velocity's deviation from Keplerian, radial velocity $v_R$, and vertical velocity with respect to the midplane $v_z$. The solid lines are from the top surface (i=45 deg), and the dashed lines are from the bottom surface (i=-45 deg). Note $v_z>0$ refers to velocity vectors pointing away from the midplane in all cases.}
    \label{fig:3v}
\end{figure}

We notice a quasi-axisymmetric ring of redshifted gas near the gap center, which is not explainable by radial motion (accretion). To understand this feature better, we find the best fit of the axisymmetric value of $v_\phi$, $v_R$, and $v_z$ for the emission layer at each radius. The results are plotted in Figure~\ref{fig:3v}. We also performed the same analysis on the other side of the inclined disk ($i=-45$ deg), with the results plotted as dashed lines. The gas surface density profile at the same snapshot is in the top panel for reference. Outside the gap region, there's a shallow gap between 16 and 20 au and another shallow gap with an adjacent density bump at 4-6 au; they appear to be a consequence of the fast accretion caused by the concentrated poloidal magnetic field in the planet-opened gap, which removes mass from outside the gap to the inside of it. In contrast, the secondary inner gap in \citet{2023ApJ...946....5A} is much deeper with a strong magnetic flux concentration.
The azimuthal velocity profiles are almost identical on both sides of the disk at all three $z_0$ locations, except the $z_0=0.3$ where anywhere within 5 au would be dominated by the disk wind. For $z_0=0.3$, a larger deviation from the other two $z_0$ locations is also expected in the other two velocity components. Nevertheless, the azimuthal velocity kink is still a solid method to uncover a planetary perturbed in an MHD wind-launching disk. 
The radial velocity is about 50 m/s between 12 and 20 au for all cases but shows significant fluctuations in the inner disk. A feature of surface accretion driven by MHD wind is the narrow transition layer of $v_R$ near the disk surface. Within one scale height, $v_R$ could change its sign entirely (Figure~\ref{fig:azcutvrcs}a). 
But this could not explain the large $v_R$ deviation in the gap for t=100~\tp. Take the middle case $z_0=2$ as an example, from Figure~\ref{fig:fidstream900}, we can see that the fast accretion layer completely encapsulates the bottom emission surface, while the top surface probes the large vortex just below the wind base. The same break of top-bottom symmetry is also shown in the vertical velocity. The bottom emission surface has a relatively fast flow moving away from the midplane at the planet-opened gap (see the dashed lines in Figure~\ref{fig:3v}), which shows a remarkable resemblance to the flow pattern inferred from observations of the HD163296 and AS209 disks. In our case, it is caused by the accretion layer being displaced from the midplane by a large vortex, which also makes the $v_z$ in the top emission layer jump between positive and negative values across the gap. The negative $v_z$ near the planet's orbit corresponds to the part of the meridional vortex that moves towards the midplane, which gives rise to the quasi-axisymmetric redshifted ring near the gap center noted earlier in the LOS velocity map (see Figure~\ref{fig:vlos}).

\begin{figure*}
    \centering
    \includegraphics[width=1.0\textwidth]{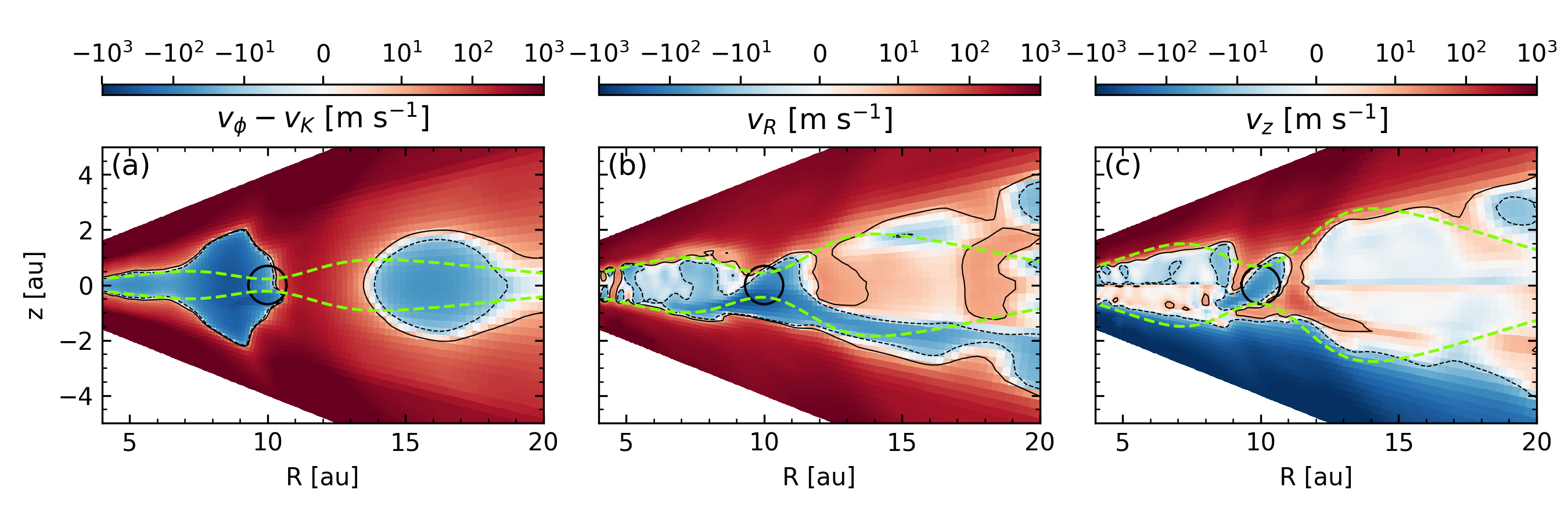}
    \caption{Azimuthally averaged map of three velocity components: (a) $v_\phi-v_K$, (b) $v_R$, and (c) $v_z$. All panels use the same ``SymLog'' scale with the linear threshold set to $\pm10~{\rm m~s^{-1}}$, and the solid and dashed contours are values of $10~{\rm m~s^{-1}}$ and $-10~{\rm m~s^{-1}}$, respectively. From left to right, the dashed green lines in the three panels are ``emission surfaces'' with $z_0$ = 0.1, 0.2, and 0.3, respectively.}
    \label{fig:3vmap}
\end{figure*}

To comprehensively demonstrate the spatial distribution differences among the three velocity components, we show in Figure~\ref{fig:3vmap} their two-dimensional R-z color maps at 
t = 100~\tp, using azimuthal averaging similar to Figure~\ref{fig:fidp6ring}. As shown in Figure~\ref{fig:3v}b, $v_\phi$ maintains mirror symmetry at different heights. The widespread sub-Keplerian region between 14-20 au is likely caused by the overpressure at the outer gap edge, resulting in a greater deviation from Keplerian motion than a disk with a smooth density profile. The distribution of $v_R$
  shows a particularly clear break in mirror symmetry across a broad region from radii 5-20~au, where the main accretion flow is near the bottom disk surface, only partially extending to the upper disk surface in the gap due to a vortex. Near the midplane, both the negative $v_R$ region inside the planet's orbit and the positive $v_R$
  region outside the orbit extend significantly in radius. The distribution of $v_z$
  also directly reveals the impact of the asymmetric accretion layer. The distribution of $v_z$ is also asymmetric relative to the midplane, as indicated by the gap's large red triangle-shaped region to the planet's right side,  and blue region across the Hill sphere. It results from the combined effects of the accretion layer and the vortex. The $v_z$ distribution shows the general trend of gas movement in the accretion layer at the bottom surface: it first approaches the midplane from the outer gap edge (positive $v_z$). Then it moves back towards the surface. It is consistent with the streamlines shown in Figure~\ref{fig:fidstream900}. 
\begin{figure*}
    \centering
    \includegraphics[width=1.0\textwidth]{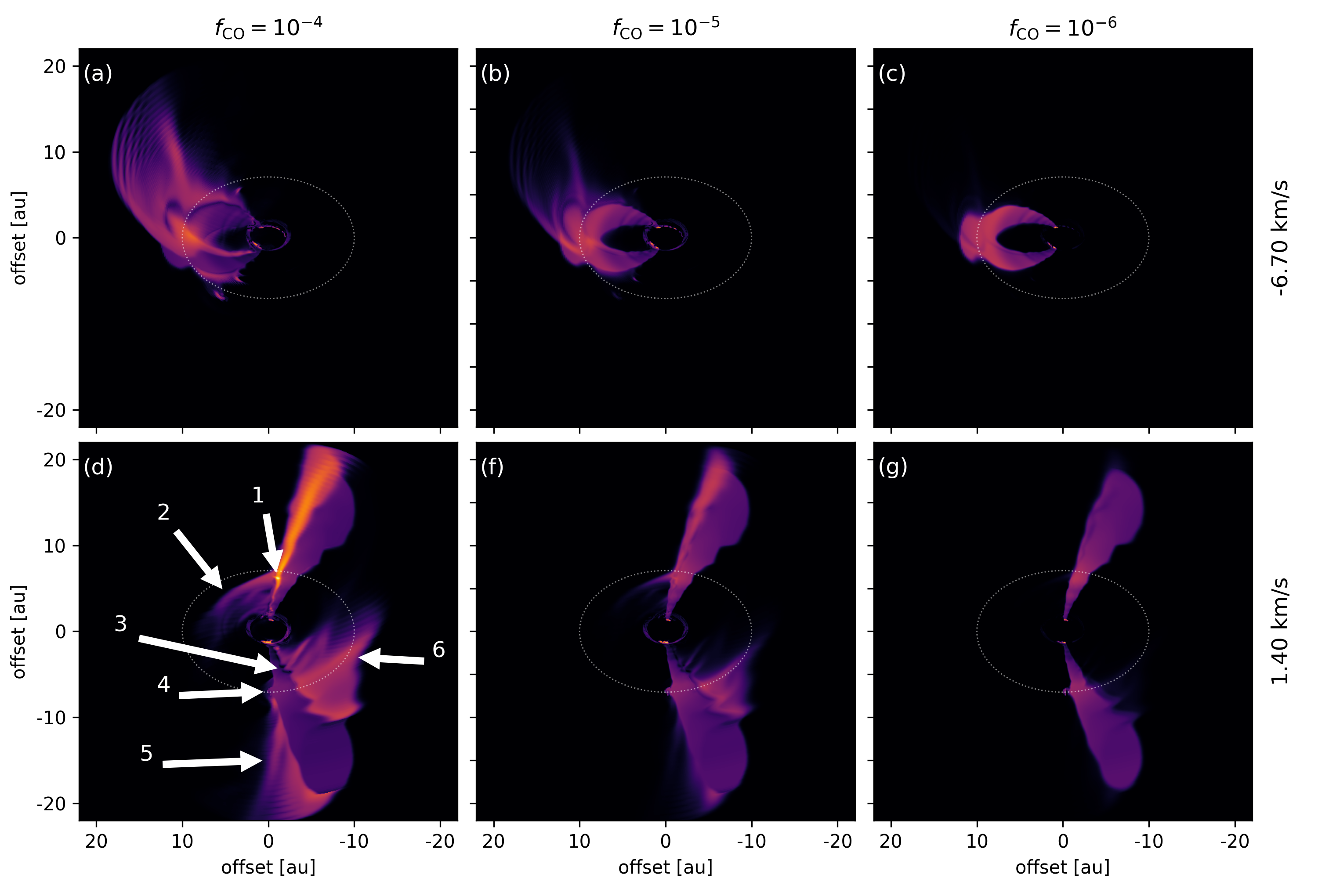}
    \caption{Synthetic line emission maps of $^{12}$CO $J=2-1$ at two different velocity channels. The top row has a velocity of -6.7 km/s, and the bottom row is 1.4 km/s. From left to right, each column is assigned with a different CO abundance, $10^{-4}$ (ISM value), $10^{-5}$, and $10^{-6}$. The thin dotted ellipse in each panel indicates the planet's orbit. We highlighted six features in panel (d).}
    \label{fig:channel}
\end{figure*}

We generated the $^{12}$CO $J=2-1$ emission map of the wind-driven accretion disk using~\texttt{RADMC-3D} \citep{2012ascl.soft02015D} assuming a disk inclination angle of $45^\circ$. We only included the disk region with $r<23$ au where the midplane is mesh-refined by at least 2 levels. The temperature structure is the same as used in the MHD simulation. We annotated six features in panel (d) of Figure~\ref{fig:channel}: (1) the bright spot is the disk wind that is denser than its surroundings, probably from the inner spiral density wake, and there are similar bright spots in panel (a), elevated from both top and bottom disk surfaces; (2) disk wind from the bottom disk surface has a low LOS velocity because the $v_r$ has a positive contribution to $v_{\rm LOS}$ while the super Keplerian $v_{\rm \phi}$ (counter-clockwise rotation) is negative on the LOS; (3) the velocity kink caused by the gap opening planet at the top emission surface; (4) the tiny bulge is the from the circumplanetary disk; (5) and (6) are disk winds from bottom and top disk surface, and both of them have a redshifted $v_{\phi}$ and a blueshifted $v_{\rm r}$ partially canceling out each other along the LOS. The velocity kinks on the disk surface exhibit structures similar to those in hydrodynamic simulations \citep{2021A&A...650A.179I}, aligning with the $v_\phi$ profile seen in Figure~\ref{fig:3v}b. The planetary torque predominantly shapes the morphology of $v_\phi$ around the gap. Kinks also appear in the outer regions, originating from a shallow gap at 18 AU (see Figure~\ref{fig:3v}a). When assuming an ISM level ($f_{\rm CO}=10^{-4}$) CO abundance, the disk wind extends noticeably beyond the upper and lower surfaces of the disk. However, an abundance one order of magnitude lower shows wind primarily at the disk's outer part. The wind becomes nearly invisible if the CO abundance is two orders of magnitude lower.

The current ALMA molecular line disk observations typically do not show obvious wind signatures. There are several possible reasons for the non-detection. First, our model has a relatively strong wind driven by a disk with a high accretion rate of 
 $\sim 10^{-6}m_\odot~{\rm yr^{-1}}$, which is higher than most of the relatively evolved disks with high resolution and sensitivity line observations \citep[e.g., AS209][]{2021ApJS..257....5Z}. Second, the CO abundance in disks (and the winds from them) might be one to two orders of magnitude lower than ISM values \citep[e.g.,][]{2021ApJS..257....5Z}, which may make it harder to detect the wind in CO; more optically thick lines, such as those from neutral carbon may be required to detect the wind. 3) The current molecular line observations typically probe disk regions larger than shown in Figure~\ref{fig:channel}, where the wind could be weaker. For example, \citet{2023NatAs...7..905F} suggested that the MHD wind from the inner disk can block high-energy radiation from the central star, thus inhibiting the generation of photoevaporation wind in the outer disk. Disk accretion may also concentrate the poloidal magnetic flux in the inner disk, leaving the outer disk too weakly magnetized to launch a significant wind. 

\subsection{Gas dynamics around the planet}
\label{sec:planet}

\begin{figure*}
    \centering
    \includegraphics[width=1.0\textwidth]{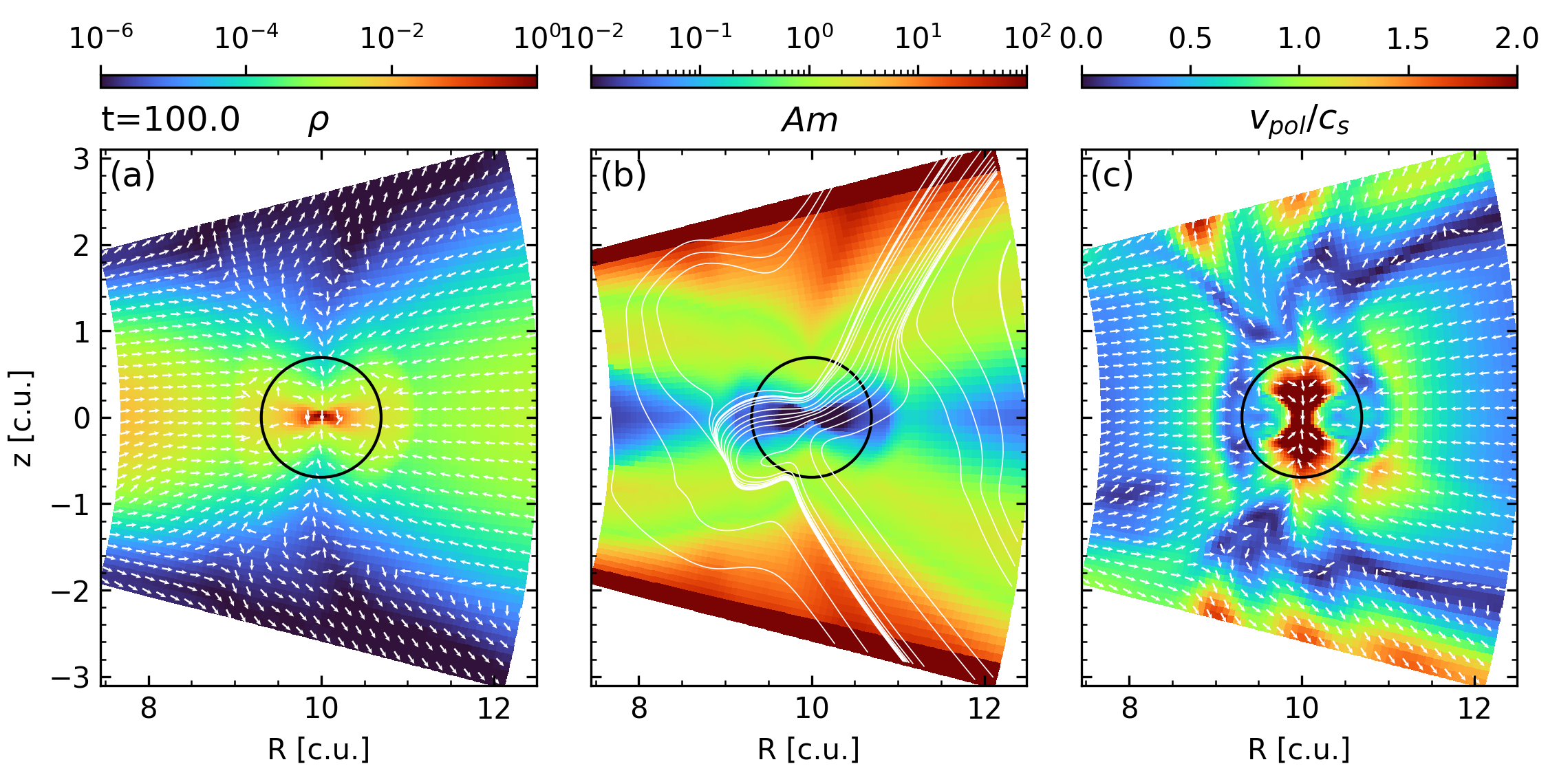}
    \caption{Azimuthally averaged gas properties around the planet, within the 8 cells along $\phi$ direction. Panel (a): gas density (in code units) with white arrows representing velocity vectors in the poloidal plane; (b): ambipolar Elsasser number and poloidal magnetic fields (solid white lines); (c): poloidal velocity divided by the local sound speed with white arrows representing velocity vectors in the poloidal plane. The solid black circle in each panel shows the planet's Hill sphere. See the supplementary online material for an animated version of this figure.}
    \label{fig:fidp3pl}
\end{figure*}

In Figure~\ref{fig:fidp3pl} we plotted the gaseous and magnetic properties near the planet at the snapshot at t=100\tp. To focus on the area that's most perturbed by the planet, all panels are azimuthally averaged within 4 cells on each side of the planet. In panel (a), the dense circumplanetary disk (CPD hereafter) extends 16 cells in diameter, which is about half the size of the Hill sphere. The small size of the CPD means its kinematics signature is limited to a tiny spatial scale that makes direct observations extremely challenging. The signature of CPD is seen between $\pm1.4$ km/s channels in our synthetic observations with three different CO abundances (Figure~\ref{fig:channel}), which reflects its large velocity dispersion in the tiny Keplerian disk. But the CPD's little ``bulge'' in the channel map could be easily lost after beam convolution with background noise, especially when it is close to the PPD's bright Keplerian pattern. The CPD candidate in AS209 appeared as a point source in the $^{13}$CO map but not the more abundant $^{12}$CO, and only in a very narrow velocity (channel) range \citep{2022ApJ...934L..20B}. The area that produces large LOS velocity is only a small portion of the CPD, further limiting the kinematic confirmation of a Keplerian CPD through observations. In summary, molecules with less abundances may be preferred for CPD detection as they are less likely to be ``overshadowed'' by the emission from the PPD. On the other hand, the less abundant molecules like C$^{18}$O, HCN, and C$_2$H are usually detected with narrower spatial distributions that do not necessarily cover the CPD location \citep{2023ApJ...950..147G}, and/or suffer from a low signal-to-noise ratio \citep{2021ApJS..257....4L}. Improvement in both spatial and spectrum resolution could better isolate the CPD signature, and a longer integration time is needed for future observations of CPD candidates. Inside the Hill sphere, gas accretes onto the CPD from both the south and north hemispheres and leaves the CPD through its equatorial plane. The decretion flow collides with the incoming gas from both sides of the gap, forming two lobes that isolate the CPD from the gap edges. Moving along the two poles away from the CPD, the infall towards the plane eventually transitions to an outflow. Notably, there is a puffed-up layer starting from R=9 c.u. separated from the inner gap edge, resulting from the enhanced density of the inner spiral density wake. This structure is truncated by the accretion flow directly above the planet when it approaches the planet's orbit. 
The magnetic field lines have two major features: strong concentration on the CPD and significant asymmetry with respect to the midplane.
Because of the CPD's high density, the strong diffusion effectively decouples the magnetic field lines from the accretion and rotation motions inside it. However, the look-up table may overestimate the degree of decoupling (See Section~\ref{sec:discussion} for a further discussion). The poloidal velocity color map saturates near the planet's polar regions in panel (c), as our local isothermal assumption supports a runaway accretion of the massive planet. Other sonic regions include the edge of two spiral-density waves and the lower wind base in the gap. 

\section{Discussion and Conclusion}
\label{sec:discussion}

Our high-resolution 3D simulations reveal that magnetic fields significantly alter the development and morphology of the gap created by a protoplanet. Specifically, we observe that the embedded planet can lead to a concentration of the poloidal magnetic flux around its orbit, thereby enhancing angular momentum removal in the gap region. This process results in deeper and more pronounced gap formation, akin to those observed in inviscid disk models (e.g., gap depth versus time in Figure~\ref{fig:hillmass}b), and is characterized by a fast, trans-sonic accretion layer that is asymmetrically distributed with respect to the disk midplane. Notably, the accretion layer makes both updraft and collapsing flow possible in the gap. While a disk wind may not directly limit the growth of the planet, wind-driven accretion can indeed more rapidly deplete the mass reservoir feeding the planet. This suggests that protoplanetary disks may exhibit more complex gas kinematics potentially accessible through observations: beyond the non-axisymmetric features dominated by planets, magnetic fields cause both radial and vertical motions of the gas to extend beyond the vicinity of the planet, and the directions of these movements are no longer solely dependent on gravitational forces and pure hydrodynamic effects like the pressure gradient. 

Comparing our findings with those of \citet{2023ApJ...946....5A} and \citet{2023A&A...677A..70W}, we note a similar emphasis on the significant role of magnetized disk winds and the concentration of the poloidal magnetic flux in the gap. Among the three asymmetries of gap layout in \citet{2023A&A...677A..70W}, we did not find notable radial width asymmetry or radial depth asymmetry. In the azimuthal direction, our results also demonstrate asymmetries that, while not drastic, are still relatively significant. During the early stages of gap opening, there is a relatively higher surface density on the side of the planet's leading edge, which may be associated with the lower magnetic field line density observed in this region as illustrated in Figure~\ref{fig:azcutvrcs}; a weaker magnetic field on this side reduces the accretion rate. After t=80~\tp, the high-density region shifts to the trailing edge side. By t=90~\tp, this area shrinks to within $\pi/4$. This region remains stable until t=110~\tp, after which there is a tendency for it to shift back towards the leading edge side. 

Our work distinguishes itself by utilizing the magnetic diffusion lookup table from \citet{2023MNRAS.523.4883H}, allowing for a more detailed and accurate depiction of magnetic diffusion processes within the disk. This approach enables us to capture the complex feedback mechanisms between disk magnetization, wind properties, and planet-induced structures that were not fully addressed by the fixed ambipolar Elsasser number (Am) profiles used in the aforementioned studies. Several features are unique in our simulations due to this diffusion profile. 

First, spontaneous symmetry breaking could stem from the vertical stratification of magnetic diffusion inside the disk. In tests without a planet, we found that asymmetry begins to emerge after just 20~\tp, even leading to scenarios where disk winds occur on only one side. The planet torque is a perturber with mirror symmetry to the midplane, thus it could potentially help the disk maintain symmetry. The planet-free spontaneous symmetry breaking has also been confirmed by linear theory \citep{2024ApJ...972..142W}. The one-sided accretion layer in our simulation is already formed in the outer disk. This is different from the accretion stream in \citet{2023A&A...677A..70W} (see their Fig.7 and Fig.A.1), where it enters the gap through the midplane and is only slightly biased towards the bottom of the disk. This is because a constant lower Elsasser number ($Am=1$) was used inside the disk, leading to accretion dominated by the densest midplane. In addition, the fixed $Am$ in \citet{2023A&A...677A..70W} is not completely unchanged because it has a lower limit that depends on the ratio of Alfv\'en velocity and sound speed ($v_a/c_s$), which, in the gap, due to reduced density and increased magnetic field strength, locally decreases the diffusion strength, with a corresponding increase in $Am$. 

Second, the stronger magnetic diffusion outside the gap in our simulation maintains a more laminar disk. For example, in Model $3Mj\beta_3$ (3 Jupiter-mass planet, with disk's initial midplane plasma $\beta=10^3$) of \citet{2023A&A...677A..70W}, the flow field within the disk exhibits a high degree of temporal variability, with the position and shape of large vortices changing significantly within one or two local orbits. In contrast, our accretion stream and vortex remained stable longer (greater than 20). The MRI active disk in \citet{2023ApJ...946....5A} is also highly dynamic and produces planet-free magnetic flux concentration that drives a second gap inside the planetary orbit.

Although the look-up table is a step forward over fixed diffusion profiles, several caveats are worth mentioning. As mentioned in Section~\ref{sec:planet}, the look-up table may be unreliable in high-density regions such as the CPD. This limitation arises not only from the CPD's density exceeding initial values but also from inaccuracies in ionization estimates. The second parameter, vertical location, aims to capture the complex effects of varied ionization sources. Near the midplane, ionization is dominated by diffuse sources, including down-scattered X-ray photons, cosmic rays, and short-lived radionuclides. Furthermore, hard X-ray photons from the central star may penetrate 1-2 scale heights below the disk surface. In the original 2D simulation, diffuse ionization attenuation was calculated using column density derived from the local density and scale height. Given the CPD's smaller scale height compared to the protoplanetary disk (PPD), this may lead to an overestimated attenuation. 
Apart from these microphysics processes, the gap morphology in the 2D simulation differs somewhat from 3D. In particular, the 2D simulation does not capture non-axisymmetric structures like spiral shocks and the CPD. The look-up table is constructed from the diffusion profile at the gap center, encompassing the widest density range. However, the behavior at the gap edge may differ. Future improvements should incorporate radial variations in the look-up table to enhance its reliability across the gap region.

The fast accretion layer across the gap is reminiscent of the dust conveyor belt in \citet{2022MNRAS.516.2006H}. The layer represents a stable gas feature resulting from efficient magnetic braking associated with the toroidal magnetic field reversal. It serves a dual role as both a trap and a conveyor for dust grains. These grains rapidly migrate inward, driven by gas advection and their inherent radial drift. Interestingly, the dust grains do not congregate at the density peaks within the gas rings, a scenario one might expect if the pressure-gradient driven radial drift were the dominant factor in dust aggregation. Rather, they are primarily found in the inner portions of these rings, where the gas rotation velocity surpasses the local Keplerian speed. This pattern suggests that the dust dynamics are more significantly impacted by the overall gas motions than by the pressure gradients. Similar meridional circulation patterns also occur in the vicinity of the planet-opened gap and the outer gap edge. The accretion layer is expected to similarly transport dust grains across the pressure maximum at the gap edge. However, since the accretion layer is concentrated on one side of the midplane, it may be less efficient in transporting settled larger grains than smaller grains. On the other hand, both the planet and magnetic fields induce significant vertical motions within the disk, facilitating the lifting of more grains from the midplane into the accretion layer. Adding the third dimension, the ``shrunk'' horseshoe orbit forms a vortex between the planet and the trailing Lagrangian point L5. The overdensity is located in the same area but can slowly shift to the other side of the planet. Dust clumps near L4 and L5 have been observed in LkCa 15 disk \citep{2022ApJ...937L...1L}. Note the overdensity is seen at the leading side of the planet in hydro simulations with wind torque \citep{2020A&A...633A...4K}, and the MHD disk in \citet{2023A&A...677A..70W} showed a similar oscillation of mass distribution along the azimuthal dimension. This indicates that under the combined influence of magnetic fields, disk wind, and accretion streams, the low-density horseshoe region could exhibit unprecedentedly complex behavior. It would be interesting to explore the dust dynamics and distribution near planet-opened gaps in non-ideal MHD disks quantitatively in the future. 

In summary, we have carried out a 3D simulation of the gap opening by an embedded planet in a non-ideal MHD disk using magnetic diffusivities from a published axisymmetric simulation with consistent thermochemistry. The main results are as follows:

\begin{itemize}[leftmargin=5pt,labelwidth=0pt,itemindent=!]
    \item We find a strong concentration of the poloidal magnetic flux in the planet-opened gap (see Figures~\ref{fig:sigevo}b and \ref{fig:fidp6ring}b), in agreement with previous work. In our simulation, the mass depletion by the embedded planet makes the low-density gap region better coupled to the magnetic field than the bulk of the disk material outside the gap (see Figure~\ref{fig:fidp6ring}b). 
    
    \item The relatively strong magnetic field and good field-matter coupling enable efficient magnetic braking of the gap material, driving a fast accretion layer that is significantly displaced from the disk midplane (see Figure~\ref{fig:fidp6ring}e). The accretion layer is located where the field lines bend most in the radial (see Figure~\ref{fig:fidp6ring}b) and azimuthal (see Figure~\ref{fig:azcutvrcs}b) directions. Its angular momentum removal is dominated by the $z\phi$-component of the Maxwell stress (see Figure~\ref{fig:stress}).  
    
    \item The magnetically driven fast accretion helps deplete the gap at late times compared to the non-magnetic (hydro) case (see Figure~\ref{fig:hillmass}b). This magnetically driven mass depletion reduces the mass reservoir in the circumplanetary environment and is expected to negatively impact the growth of the planet (see Figure~\ref{fig:hillmass}a). In the fast accretion layer, the region of horseshoe orbits on the trailing side of the planet is greatly reduced (see Figure~\ref{fig:horseshoe}), which is expected to impact the corotation torque on the planet.   

    \item The poloidal gas kinematics in the gap is dominated by an interplay between the magnetically driven disk wind and magnetically driven fast accretion layer that, in our simulation, is displaced below the midplane. Sandwiched between the disk wind from the top surface and the fast accretion layer below the midplane is a large, persistent vortex that dominates the gas kinematics in the meridional plane (see Figure~\ref{fig:fidstream900}). The meridional vortex in the gap may be observable as a redshifted (or blueshifted, depending on the disk inclination) ring (see Figure~\ref{fig:vlos}). The fast accretion layer may also be observable if located close enough to the surface. 
\end{itemize}

\section*{Acknowledgements}
We thank the referee for a detailed constructive report that improved the presentation of the paper. ZYL is supported in part by NASA 80NSSC20K0533 and NSF AST-2307199. Our simulations are made possible by an XSEDE allocation (AST200032). Z.Z. acknowledges support from the National Science Foundation under CAREER Grant Number AST-1753168 and support from NASA award 80NSSC22K1413. Figures in this paper were made with the help of \verb|Matplotlib| \citep{Hunter2007} and
\verb|NumPy| \citep{2020Natur.585..357H}. \verb|cblind| is also used in one of the colormaps.

\section*{Data Availability}

The data from the simulations will be shared on reasonable request to the corresponding authors.

\bibliographystyle{mnras}
\bibliography{ref_mhd_dust}


\bsp	
\label{lastpage}
\end{document}